\documentclass[twocolumn,showpacs,preprintnumbers,amsmath,amssymb]{revtex4}

\begin{document}

\title{Supersymmetric string model with $30$ $\kappa$--symmetries
in an extended $D=11$ superspace and ${30\over 32}$ BPS states}
\author{Igor A. Bandos$^{\dagger\ast}$, Jos\'e A. de
Azc\'arraga$^{\dagger}$, Mois\'es Pic\'on$^{\dagger}$ and Oscar
Varela$^{\dagger}$} \affiliation{$^{\dagger}$Departamento de
F\'{\i}sica Te\'orica and IFIC (CSIC-UVEG), 46100-Burjassot
(Valencia), Spain
\\
$^{\ast}$Institute for Theoretical Physics, NSC KIPT, UA61108,
Kharkov, Ukraine}

\begin{abstract}
A supersymmetric string model in the $D=11$ superspace maximally extended by
antisymmetric tensor bosonic coordinates, $\Sigma^{(528|32)}$, is
proposed. It possesses $30$ $\kappa$-symmetries and $32$ target
space supersymmetries. The usual preserved
supersymmetry--$\kappa$-symmetry correspondence suggests that it
describes the excitations of a BPS state preserving all but two
supersymmetries. The model can also be formulated in any
$\Sigma^{({n(n+1)\over 2}|n)}$ superspace, $n=32$ corresponding to
$D=11$. It may also be treated as a `higher--spin generalization'
of the usual Green--Schwarz superstring. Although the global
symmetry of the model is a generalization of the super--Poincar\'e
group, ${\Sigma}^{({n(n+1)\over 2}|n)}\times\!\!\!\!\!\!\supset
Sp(n)$, it may be formulated in terms of constrained $OSp(2n|1)$
orthosymplectic supertwistors. We work out this supertwistor
realization and its Hamiltonian dynamics.

We also give the supersymmetric $p$--brane generalization of the model. 
In
particular, the $\Sigma^{(528|32)}$ supersymmetric membrane model describes
excitations of a ${30\over 32}$ BPS state, as the
$\Sigma^{(528|32)}$ supersymmetric string does, while the supersymmetric 
3--brane and 5--brane correspond, respectively, to ${28\over
32}$ and ${24\over 32}$ BPS states. 
\end{abstract}


\maketitle

\narrowtext

\section{Models in non-standard superspaces:
particles, strings, BPS preons and higher spin theories.}

In the early period of superstring theory, when it was found that
all  $D=10$ supergravities appear as low energy limits of
superstring models, a question arose: what is the origin of
maximally extended $D=11$ supergravity? Its relation with the
supermembrane \cite{BST} was established by studying the
supermembrane action in a supergravity background; however, a
straightforward quantization of the supermembrane was fraught with
difficulties. An indication was found \cite{Hoppe} that the
quantum state spectrum of the supermembrane is continuous, a
problem now sorted out by treating \cite{Nicolai} the
supermembrane as an object composed of D$0$--branes in the
framework of the Matrix model approach \cite{(M)atrix}. Another
aspect of the same problem was that the membrane was shown to
develop string--like instabilities \cite{Hoppe}. The
Green--Schwarz superstring is free from these problems, but it is
a $D=10$ theory. Thus, it was tempting to search for possible new
classical $D=11$ superstring models hoping that, after
quantization, their low energy limit would be $D=11$ supergravity.
Such a search requires, clearly, going beyond the standard
superspace framework: in moving from $D=10$ to $D=11$ one 
has to add also extra bosonic degrees of freedom, thus arriving
to an {\it enlarged} $D=11$ superspace rather than to
the standard one. 

\subsection{Curtright supersymmetric string model in the
enlarged  $D=11$ superspace $\Sigma^{(528|32)}$}

A first example of a supersymmetric string action in an enlarged 
$D=11$ superspace was found in \cite{Curtright}. The model, 
possessing $32$ supersymmetries and
$16$ $\kappa$--symmetries, was constructed in the enlarged
superspace $\Sigma^{(528|32)}$. This contains $32$ fermionic
coordinates $\theta^\alpha$ and $528$ bosonic coordinates $x^\mu ,
y^{\mu\nu}, y^{\mu_1\ldots \mu_5}$ ($y^{\mu\nu}=- y^{\nu\mu}\equiv
y^{[\mu\nu]}$, $y^{\mu_1\ldots \mu_5}=y^{[\mu_1\ldots \mu_5]}$)
which may be collected in a symmetric spin-tensor
$X^{\alpha\beta}= X^{\beta\alpha}$,
\begin{equation}\label{11+}
X^{\alpha\beta}= {1\over 32} x^\mu \Gamma_\mu^{\alpha\beta} -
{1\over 2!\, 32} y^{\mu\nu} \Gamma_{\mu\nu}^{\alpha\beta} +
{1\over 5! \, 32} y^{\mu_1\ldots \mu_5} \Gamma_{\mu_1\ldots
\mu_5}^{\alpha\beta} \; ,
\end{equation}
so that the coordinates of $\Sigma^{(528|32)}$ are
\begin{eqnarray}
\label{528} & {\cal Z}^{{\cal M}}= (X^{\alpha\beta},
\theta^\alpha)\; , \quad X^{\alpha\beta}= X^{\beta\alpha}\; ,
\qquad  \\ & \nonumber \alpha, \beta = 1,2, \ldots , 32 \; .
\end{eqnarray}

Due to the special properties of the eleven--dimensional gamma
matrices, the model of \cite{Curtright} may also be
restricted to the superspaces $\Sigma^{(66|32)}$ [$(x^\mu,
y^{\mu\nu}, \theta^\alpha)$, with 66 bosonic coordinates] and
$\Sigma^{(462|32)}$ [$(x^\mu, y^{\mu_1\ldots \mu_5},
\theta^\alpha)$, with $462$ bosonic coordinates] \footnote{All
these superspaces $\Sigma^{(528|32)}$, $\Sigma^{(462|32)}$ and
$\Sigma^{(66|32)}$, considered as supergroup manifolds, may be
seen as central extensions of an abelian $32$ dimensional
fermionic group  by tensorial (Eq.~(\ref{11+})) bosonic groups
\cite{JdA00}.}. For the sake of definiteness, we shall call here
{\it maximal superspaces} to those with bosonic coordinates of
symmetric `spin--tensorial' type, like $\Sigma^{(528|32)}$ and its
counterparts $\Sigma^{({n(n+1)\over 2}|n)}$,
\begin{eqnarray} \label{Sn}
& \Sigma^{({n(n+1)\over 2}|n)}\; =   \; \left\{{\cal Z}^{\Sigma}=
(X^{\alpha\beta}, \theta^\alpha)\right\} \; , \quad
X^{\alpha\beta}= X^{\beta\alpha}\; , \qquad \\ \nonumber & \alpha,
\beta = 1,2, \ldots n \quad ,
\end{eqnarray}
where $n=2^l$. This name distinguishes the
 $\Sigma^{({n(n+1)\over 2}|n)}$
superspaces from other, not maximally (in the bosonic sector)
extended superspaces like $\Sigma^{(66|32)}$ and
$\Sigma^{(462|32)}$ whose bosonic coordinates may be described by
a spin--tensor $X^{\alpha\beta}$ only if it satisfies some
conditions.

The $\Sigma^{(528|32)}$ superspace has a special interest because
it is the supergroup manifold associated with the maximal $D=11$
supersymmetry algebra \cite{FRE,HP82,M-alg}
\begin{eqnarray}\label{QQP}
& \{ Q_\alpha , Q_\beta\} = P_{\alpha \beta} \; , \quad P_{\alpha
\beta}= P_{\beta\alpha} \; , \quad [Q_\alpha , P_{\beta\gamma}]=0
\; ,\quad
\\
& \alpha , \beta , \gamma = 1,2, \ldots 32 \quad , \nonumber\\
\label{n32} & P_{\alpha\beta} = P_\mu \Gamma^\mu_{\alpha\beta} +
Z_{\mu\nu} \Gamma^{\mu\nu}_{\alpha\beta} +  Z_{\mu_1\ldots \mu_5}
\Gamma^{\mu_1\ldots \mu_5}_{\alpha\beta}\; ,
\end{eqnarray}
called M--theory superalgebra or M--algebra \cite{M-alg}
\footnote{See \cite{gM-alg,JdA00} and refs. therein for further
generalizations of the M--theory superalgebra and for their
structure.}. This algebra encodes a full information about the
nonperturbative BPS states of the hypothetical underlying
M--theory: as it was shown in \cite{JdAT}, the additional bosonic
generators $Z_{\mu\nu} = - Z_{\nu\mu}=Z_{[\mu\nu]}$,
$Z_{\mu_1\ldots \mu_5}=Z_{[\mu_1\ldots \mu_5]}$ of the M--algebra
(\ref{QQP}) are related to the topological charges of the
supermembrane and the super--M5--brane \footnote{This result was
extended in \cite{Hull97} by showing that these generators  also
contain a contribution from the topological charges of the
eleven--dimensional Kaluza--Klein monopole ($Z_{0\mu_1\ldots
\mu_4}\propto \epsilon_{0\mu_1\ldots \mu_4 \nu_1\ldots \nu_6}
\tilde{Z}^{\nu_1\ldots \nu_6}$) and of the M9--brane ($Z_{0\mu
}\propto \epsilon_{0\mu \nu_1\ldots \nu_9} \tilde{Z}^{\nu_1\ldots
\nu_9}$) which is usually identified with the Ho\v rava--Witten
hyperplane \cite{HW96} (for the Kaluza--Klein monopole and the M9
brane only bosonic actions are known
\cite{BergshoeffKK,BergshoeffM9}).}. These `one--brane' BPS states
can be associated  with solitonic solutions \cite{Duff,Stelle} of
the `usual' $D=11$ supergravity \cite{CJS} or with fundamental
M-theory objects described by their worldvolume actions
\cite{BST,blnpst}.

Note that although the M--algebra (\ref{QQP}) leads naturally to a
$D=11$ interpretation when the splitting (\ref{n32}) is used, it
also allows for a $D=10$ type IIB treatment when one considers the
$\alpha=1,...,32$ index as a double one $\alpha= \check{\alpha}
I$, where $\check{\alpha}$ labels the components of a $D=10$
Majorana--Weyl spinor, $\check{\alpha}=1,\ldots 16$, and $I$ is an
internal index, $I=1,2$. Then one uses the direct product of
$16\times 16$ (Majorana--Weyl) ten--dimensional sigma matrices,
$\sigma^\mu_{\check{\alpha}\check{\beta}}$ (or
$\tilde{\sigma}_\mu^{\check{\alpha}\check{\beta}}$) and real
$2\times 2$ matrices $\delta^{IJ}, \tau_1^{IJ}, i\tau_2^{IJ},
\tau_3^{IJ}$ to write a $D=10$ type IIB counterpart of Eq.
(\ref{n32}) \cite{Bars97,M-alg}. Similarly, a $D=10$ type IIA
treatment is also possible as the $D=10$ gamma matrices coincide
with the $D=11$ ones. As a result, the information about
nonperturbative BPS states of the $D=10$ superstring theories
(including Dirichlet superbranes) can also be
 extracted from the algebra (\ref{QQP}). Moreover,
it encodes as well all the duality relations between different
$D=10$ and $D=11$ superbranes. These facts add further reasons to
call (\ref{QQP}) the M--theory superalgebra \cite{M-alg}.

\subsection{Maximal $\Sigma^{(528|32)}$ superspace, BPS preons
and other BPS states with supernumerary supersymmetries}

Interestingly enough, an algebraic classification of all BPS
states may be achieved by introducing \cite{BPS01} the
hypothetical basic constituents of M--theory. These are the {\it
BPS preons} $|\lambda>$, which are characterized by the relation
\begin{eqnarray}\label{BPSdef}
\hat{P}_{\alpha\beta} |\lambda > = \lambda_\alpha\lambda_\beta
|\lambda > \; ,
\end{eqnarray}
where $\lambda_\alpha$ is a {\it bosonic `spinor'} ({actually, a
$GL(n,\mathbb{R})$--vector}, see footnote [81] 
and below Eq.~(\ref{susysinglets})) 
\footnote{
Note that the expression 
$p_{\alpha\beta}=\lambda_\alpha\lambda_\beta$ for the eigenvalue of the 
generalized momentum operator $P_{\alpha\beta}$ may be looked at as a 
generalization (see \cite{BL98})  
of the Penrose representation $p_{A\dot{B}}= \lambda_A 
\bar{\lambda}_{\dot{B}}$ of a $D=4$ light--like vector 
(see \cite{Pen}). Interestingly enough, its generalizations for 
tensorial charges  
($Z_{\mu\nu\rho}= \lambda\Gamma_{\mu\nu\rho}\lambda$ in $D=8=0+8$ 
and $Z_{\mu\nu}= \lambda\Gamma_{\mu\nu}\lambda$ in $D=4=1+3$
were considered, in a completely different context, in 
\cite{VST} and \cite{ZG}. 
Recently, the original $D=4$ Penrose twistor 
formalism has found an interesting application 
in the analysis of perturbative scattering amplitudes in Yang--Mills 
theories \cite{W03} which refers to {\it a string in $D=4$ twistor space} 
$CP^{3}$.}.

The states $|\lambda >$ preserve all supersymmetries {\it but one}
i.e., they are $\nu=31/32$ states. To make this transparent one
can use the $GL(32,{\mathbb R})$ automorphism symmetry
\cite{West,GGT} of the algebra (\ref{QQP}) to write the spinor
$\lambda_\alpha$ that characterizes the BPS preon state
$|\lambda>$ in the `preferred frame' $\lambda_\alpha= (1, 0 , ...,
0)$, where (\ref{BPSdef}) and (\ref{QQP}) imply
\begin{eqnarray}\label{BPSdefQ2}
(\hat{Q}_{2})^2|\lambda > = 0 \; , \quad  \dots \quad , \quad
(\hat{Q}_{32})^2|\lambda > = 0 \; .  \quad
\end{eqnarray}
For hermitian operators in a positive definite Hilbert space
Eqs.~(\ref{BPSdefQ2}) imply
\begin{eqnarray}\label{BPSdefQ1}
(\hat{Q}_{2})|\lambda > = 0 \; , \quad  \dots \quad , \quad
(\hat{Q}_{32})|\lambda > = 0 \; ,  \quad
\end{eqnarray}
which means that a BPS preon preserves all but one spacetime
supersymmetries.

BPS states $|k>$ preserving $k \ge 1$ supersymmetries can be
treated as composites of a number $\#_{p}=32-k$ of BPS preons
\cite{BPS01} (in the same way as {\it e.g.},  hadrons are composed
of quarks). Indeed, for such a state $|k>$ one can always find a
set of $32-k$ bosonic spinors $\lambda_\alpha^r$ ($r=1,...,k$)
such that
\begin{eqnarray}\label{kBPSdef}
\hat{P}_{\alpha\beta} |k> = \sum\limits_{r=1}^{\#_p=32-k}
\lambda^r_\alpha \lambda^r_\beta |k > \; ;
\end{eqnarray}
the single preon state $|\lambda >$ corresponds to $|k> \equiv
|31>$.

In this perspective, all the one--brane solutions of
11-dimensional supergravity, which preserve $16$ out of $32$
supersymmetries, correspond to composites of $16$ BPS preons.
Multibrane solutions usually preserve {\it less} than $16$
supersymmetries $(\nu<1/2)$ and thus correspond to composites of
{\it more} than $16$ preons. There also exist $pp$--wave solutions
with `supernumerary supersymmetries' \cite{CLP02,GH02,BJM02}, {\it
i.e.} with $16< k <32$. The known solutions preserving $k=18, 20,
22, 24, 26$ and $28$ supersymmetries can be considered as
composites of $\#_p =14, 12, 10, 8, 6$ and $4$ BPS preons
respectively. Initially, it seemed that solutions preserving all
supersymmetries but one, i.e. describing the excitations of a BPS
preon, could not exist in the framework of the standard brane
ansatzes used to solve the usual 11-dimensional supergravity
\cite{CJS} equations. A more general study in the context of
standard $D=11$ supergravity has shown \cite{Hull03,Duff03} that
the existence of such solutions is not ruled out. However, and
independently of whether the BPS preons can be associated with
solutions of standard supergravity or whether there is a kind of
{\it BPS preon conspiracy} preventing the existence of one BPS
preon in {\it standard} $D$=11 spacetime or superspace, BPS preons
do provide an algebraic classification of the M-theory BPS states
\cite{BPS01}. Also, dynamical models with the properties of BPS
preons are known in the $\Sigma^{(\frac{n(n+1)}2|n)}$ superspace
\cite{BL98,B02}. In this perspective such a BPS preon conspiracy,
if it exists, would rather indicate the necessity of a wider
geometric framework for a suitable description of M--theory, such
as extended superspaces and supertwistors. If, on the contrary,
solitonic solutions with the properties of BPS preons were
actually found, the extended superspaces would still provide a
useful tool for a description of M-theory \footnote{\label{footn}
There are also related reasons to consider more general
superspaces, as the ensuing {\it fields-extended superspace
democracy} associated with extended superspaces \cite{JdA00}.}.
One is led to expect that the additional tensorial coordinates of
these superspaces carry a counterpart of the information which, in
the framework of standard $D=10,11$ supergravity, is encoded in
the antisymmetric tensor gauge fields entering the supergravity
multiplets ({\it cf.}~\cite{JdA00}). This point of view may be
also supported by the observation that in the standard topological
charge treatment of the tensorial generators of the M--algebra
\cite{JdAT}, these topological charges are associated just with
these gauge fields.

The general results about the treatment of tensorial central
charges as topological charges of the corresponding branes are
certainly relevant in the more general case
 \begin{eqnarray}\label{QQPn}
& \{ Q_\alpha , Q_\beta\} = P_{\alpha \beta} \quad , \quad [
Q_\alpha , P_{\beta\gamma}] = 0 \; ,
\\
& P_{\alpha \beta}= P_{\beta\alpha}
\;,\quad\alpha,\beta,\gamma=1,2,\ldots n\quad,
\end{eqnarray}
with $n= 2^l$ for any integer $l$. The simplest representations of
the algebra (\ref{QQPn}) can be constructed on the maximal
superspace $\Sigma^{({n(n+1)\over 2}|n)}$, Eq.~(\ref{Sn}).

The $GL(n,\mathbb{R})$ symmetry of (\ref{QQPn}) becomes  broken
down to $Spin(t,D-t)\subset GL(n,\mathbb{R})$ when a
(Eq.~(\ref{n32})-like) decomposition is introduced using a
$n\times n$ realization of the gamma--matrices of a $(t, D-t)$
spacetime with $t$ timelike dimensions ($t$ is not obliged to be
one, see \cite{Bars96,Bars}). On the other side, the
$GL(n,\mathbb{R})$ symmetry is a subgroup of $Sp(2n)$, which is a
characteristic symmetry of higher spin theories.

\subsection{Models in maximal superspaces and higher spin theories}

The main problem of the approach in \cite{Curtright} is how to
treat the large number of additional bosonic degrees of freedom
\footnote{See \cite{DG} for a later related search based on an
attempt to replace the $\kappa$--symmetry requirement by a
dynamically generated projection constraint on the spinor
coordinate functions. This approach also suffers from the problem
of additional bosonic degrees of freedom.} in the coset(s)
$\Sigma^{(528|32)}/\Sigma^{(11|32)}$
($\Sigma^{(462|32)}/\Sigma^{(11|32)}$,
$\Sigma^{(66|32)}/\Sigma^{(11|32)}$), where
\begin{eqnarray}\label{11}
& \Sigma^{(11|32)} \; :\; Z^M = (x^\mu, \theta^\alpha ) \; , \quad
\mu = 0,1, \ldots , 10 \quad ,
\end{eqnarray}
is the `standard' $D=11$ superspace. Actually, one has to face
this problem in any approach dealing with enlarged superspaces
\cite{JdA00,RS,BL98,BLS99,BPS01,V01s,V01c,Manvelyan,ZL,ZU,B02}.
Thus, one has to find a mechanism that either suppresses the
additional (with respect to the usual spacetime/superspace
$\Sigma^{(D|n)}$) degrees of freedom or provides a physical
interpretation for them. In this respect $\Sigma^{({n(n+1)\over
2}|n)}$, despite having a maximal bosonic part, has some
advantages with respect to non-maximally extended superspaces (see
below (\ref{Sn})). Indeed, the bosonic sector of the maximal
superspace (\ref{Sn}),
\begin{eqnarray}\label{Sn0}
& \Sigma^{({n(n+1)\over 2}|0)}\; :  \; X^{\alpha\beta}=
X^{\beta\alpha}\; , \quad
 \alpha, \beta = 1,2, \ldots , n  \;\; ,\quad
\end{eqnarray}
was proposed for $n=4$ \cite{Fr86} as a basis for the construction
of $D=4$ higher--spin theories \cite{V01,V01s,V01c}. Moreover, it
was shown in \cite{BLS99} that the quantization of a simple
superparticle model \cite{BL98} in $\Sigma^{({n(n+1)\over 2}|n)}$
for $n=2,4,8,16$ results in a wavefunction describing a tower of
massless fields of all possible spins (helicities). Such an
infinite tower of higher spin fields allows for a nontrivial
interaction in $AdS$ spacetimes \cite{Vasiliev89,V01} \footnote{A
relation between the generalized $n=4$ superparticle wavefunctions
\cite{BLS99} and Vasiliev's `unfolded' equations for higher spin
fields was noted in \cite{V01s}. This was elaborated in detail in
\cite{Dima}, where the quantization of an $AdS$ superspace
generalization of the $n=4$ model of \cite{BL98} was also carried
out (see also \cite{Misha} for a related study of higher spin
theories in the maximal generalized $AdS_4$ superspace).}.

To give an idea of the relation between higher spin theories and
maximally extended superspaces, let us consider the  free bosonic
massless higher--spin equations proposed in \cite{V01s} (for
$n=4$). These can be collected as the following set of equations
for a scalar function $b$ on $\Sigma^{({n(n+1)\over 2}|0)}$
\begin{eqnarray}\label{hsEqb0}
\partial_{\alpha[\beta}\partial_{\gamma]\delta} b(X)=0 \; ,
\end{eqnarray}
where $\partial_{\alpha\beta}=\partial/\partial X^{\alpha\beta}$.
Eq.~(\ref{hsEqb0}) states that
$\partial_{\alpha\beta}\partial_{\gamma\delta}$ is fully symmetric
on a non-trivial solution. In the generalized momentum
representation Eq.~(\ref{hsEqb0}) reads
\begin{eqnarray}\label{hsEqbk}
k_{\alpha[\beta} k_{\gamma]\delta} b(k) =0 \; .
\end{eqnarray}
This implies that $b(k)$ has support on the
$\frac{n(n+1)}2-\frac{n(n-1)}2=n$--dimensional surface in momentum
space $\Sigma^{({n(n+1)\over 2}|0)}$ (actually, in
$\Sigma^{({n(n+1)\over 2}|0)}\backslash \{ 0\}$) on which the rank
of the matrix $ k_{\gamma\delta}$ is equal to unity \cite{V01c}.
This is the surface defined by $k_{\alpha\beta}=\lambda_{\alpha}
\lambda_{\beta}$ (or $-\lambda_{\alpha} \lambda_{\beta}$)
characterized by the $n$ components of $\lambda_\alpha$. In a
`$GL(n,\mathbb{R})$--preferred' frame (an analogue of the standard
frame for lightlike ordinary momentum),
$\lambda_\alpha=(1,0,\ldots,0)$ and the surface is the
$GL(n,\mathbb{R})$--orbit of the point
$k_{\alpha\beta}=\delta_{\alpha 1} \delta_{\beta 1}$. Thus,
Eq.~(\ref{hsEqbk}) may also be written as
\begin{eqnarray}\label{hsEqbkl}
(k_{\alpha\beta} -\lambda_{\alpha} \lambda_{\beta}) b = 0 \; ,
\end{eqnarray}
which is equivalent to writing Eq.~(\ref{hsEqb0}) in the form
\begin{eqnarray}\label{hsEqbl}
(i\partial_{\alpha\beta} -\lambda_{\alpha} \lambda_{\beta}) b = 0
\; .
\end{eqnarray}
Eqs.~(\ref{hsEqbkl}) and (\ref{hsEqbl}) may be considered as the
generalized momentum ($k_{\alpha\beta}$) and coordinate
($X^{\alpha\beta}$) representations of the definition
(\ref{BPSdef}) of a BPS preon \cite{BPS01}. The solutions of
Eqs.~(\ref{hsEqbkl}), (\ref{hsEqbl}) are the momentum and
coordinate `wavefunctions' corresponding to a BPS preon state
$|\lambda>$, $b(X)=<X|\lambda>$, $b(k)=<k|\lambda>$. These
equations also appear as a result of the quantization \cite{BLS99}
of the superparticle model in \cite{BL98} \footnote{In
\cite{V01s,Misha+} Eq.~(\ref{hsEqbl}) was written as
$(\partial_{\alpha\beta} - {\partial \over
\partial \mu^{\alpha}} {\partial \over \partial \mu^{\beta}})
b(X,\mu) = 0 $ which is an equivalent `momentum' representation
obtained by a Fourier transformation with respect to
$\lambda_\alpha$, see \cite{Dima}.}.

Thus, in contrast with other extended superspaces, the models in
the maximal superspaces $\Sigma^{({n(n+1)\over 2}|n)}$ can be
regarded as higher spin generalizations of the models in standard
superspace $\Sigma^{(D|n)}$ \footnote{Although the idea of higher
spin fields has been discussed at present for $D\leq 7$ only, the
results of \cite{BLS99} can be regarded as a first step towards
its $D=10$ generalization. Understanding the $D=11$ case is a
problem for future study.}.

\subsection{A new supersymmetric string model in $\Sigma^{(528|32)}$: 
outlook}

In Sec.~\ref{action} we present another action for a
supersymmetric string in $\Sigma^{(528|32)}$. In distinction to
the model in \cite{Curtright}, it does not use $D=11$
gamma--matrices, but instead includes two auxiliary bosonic spinor
variables, $\lambda_\alpha^+$ and $\lambda_\alpha^-$  
\footnote{\label{s-vector} Actually, the model possesses $Sp(32)$ 
symmetry besides the $SO(1,10)$ one, 
so that $\lambda_\alpha^\pm$ may be considered as symplectic vectors 
(called 's-vectors' in \cite{V01s,V01c}) rather than Lorentz spinors.
We, however, keep the `spinors' name for them keeping in mind a possibility of 
spacetime treatment, although this is not straightforward and 
requires additional study (see Sec. IC and also 
\cite{W03} for a very recent spacetime treatment of a $CP^{3}$ sigma model 
{\it i.e.}, of a string theory in twistor space, 
through its relation to Yang--Mills amplitudes).}.
As a consequence, the resulting $\Sigma^{({n(n+1)\over 2}|n)}$ 
supersymmetric string action
(although it does not include a 
Wess--Zumino term as that of
\cite{Curtright}) possesses $30$ local fermionic
$\kappa$--symmetries and provides an extended object model for a
state composed of two BPS preons (see above).

The model  can be written as well in $\Sigma^{({n(n+1)\over
2}|n)}$ for arbitrary even $n$ (although $n=2^l$ is preferable for
a spinor interpretation of the $\alpha, \beta$ indices). It
possesses $(n-2)$ $\kappa$--symmetries (Sec.~\ref{properties}).
For $n=2$, our model describes a string in the $D=3$, $N=1$
standard superspace; however, this string does not possess
any $\kappa$--symmetry ($n-2=0$) and, then, the ground state of
this string model is not a stable (BPS) state, as such a property
is guaranteed by the preservation of a non-zero number of
supersymmetries.

For $n \geq 4$ our model possesses $k>0$ $\kappa$--symmetries, $2$
for $n=4$ $(D=4)$, $6$ for $n=8$ $(D=6)$, $14$ for $n=16$ $(D=10)$
and $30$ for $n=32$ $(D=11)$, and hence describes excitations of a
two preons BPS state. Moreover, only for $n=4$ the number of
$\kappa$--symmetries is the same as that of the $D=4$ ($N=1$)
Green--Schwarz superstring. For $n \ge 8$ the number of
$\kappa$--symmetries of our model exceeds half of the number of
supersymmetries $(\nu > \frac12)$, while the $\kappa$--symmetries
of the Green--Schwarz superstring are $\frac{n}{2}$ for all
$D=3,4,6,10$ cases.

The point--like models \cite{BL98,BLS99} in maximal superspace are
enlarged superspace generalizations of the Ferber--Shirafuji
\cite{Ferber} approach to the Brink--Schwarz superparticle. The
tensionless supersymmetric string models in the maximal 
$\Sigma^{({n(n+1)\over 2}|n)}$ superspaces
\cite{ZU} \cite{B02} can be treated as generalizations of the
$D=4$ null--superstring model \cite{BZ}. In the same sense our
$\Sigma^{({n(n+1)\over 2}|n)}$ supersymmetric 
string model 
can be looked at as a generalization to the
maximal superspaces ${\Sigma}^{({n(n+1)\over 2}|n)}$ of the
Lorentz-harmonic formulation of the $D=4$ Green--Schwarz
superstring in \cite{BZstr} \footnote{The $D=4$ version of our
supersymmetric string in $\Sigma^{(10|4)}$ differs from the Lorentz harmonic
formulation \cite{BZstr} of the $D=4$, $N=1$ Green--Schwarz
superstring, by skipping the Wess--Zumino term of the latter and
by substituting $\Pi^{\alpha\beta}=dX^{\alpha\beta}- i
d\theta^{(\alpha} \theta^{\beta )}$ for $\gamma^{\alpha\beta}_\mu
\Pi^\mu := \gamma^{\alpha\beta}_\mu (dx^\mu - i d\theta\gamma^\mu
\theta)$ in the kinetic term of the action in \cite{BZstr}. The
first of the above steps clearly breaks the two
$\kappa$--symmetries of the $D=4$, $N=1$ Green--Schwarz
superstring, while the second step, which extends the bosonic body
of the standard superspace $\Sigma^{(4|4)}$ ($x^\mu \rightarrow
(x^\mu, y^{\mu\nu})$; $X^{\alpha\beta}= x^\mu
\gamma_\mu^{\alpha\beta} + y^{\mu\nu}
\gamma_{\mu\nu}^{\alpha\beta}$) to get $\Sigma^{(10|4)}$, restores
them.}.

In Sec.~\ref{Hamiltonian} we carry out a Hamiltonian analysis of
the ${\Sigma}^{({n(n+1)\over 2}|n)}$ model and describe its gauge
symmetries, including the $(n-2)$ $\kappa$--symmetries and their
`superpartners', the $(n-1)(n-2)/2$ bosonic gauge $b$--symmetries.
We also discuss there the number of degrees of freedom of our
model. In Sec.~\ref{twistor} we show that its action may be
formulated in terms of a pair of constrained $OSp(2n|1)$
supertwistors (see \cite{BL98}) which are invariant under both
$\kappa$-- and $b$--symmetries. Note that one of the constraints
imposed on the supertwistors breaks the $OSp(2n|1)$ invariance
down to the semidirect product ${\Sigma}^{({n(n+1)\over
2}|n)}\times\!\!\!\!\!\!\!\supset Sp(n)$ of the symplectic group
$Sp(n)\subset Sp(2n)$ and the supergroup associated with the
algebra (\ref{QQP}), also denoted ${\Sigma}^{({n(n+1)\over 2}|n)}$
since this superspace is the associated supergroup manifold; we
may look at ${\Sigma}^{({n(n+1)\over
2}|n)}\times\!\!\!\!\!\!\!\supset Sp(n)$ as a generalization of
the super--Poincar\'e group
${\Sigma}^{(D|n)}\times\!\!\!\!\!\!\supset SO(t,D-t)$. The
$OSp(2n|1)$ supergroup has been  considered as a generalization of
the superconformal group \cite{HP82,Fr86,BL98,Bars98,West,BPS01}
(see \cite{Bars98,West,BPS01} for the relevance of $OSp(64|1)$ in
M-theory). This generalized superconformal group symmetry is
present in massless particle-like models \cite{BL98,BLS99} and in
the tensionless superstring \cite{B02}; however, it is broken down
to ${\Sigma}^{({n(n+1)\over 2}|n)}\times\!\!\!\!\!\!\!\supset
Sp(n)$ in our tensionful supersymmetric string model (Appendix A). This is
natural: the conformal symmetry is broken in the massive
superparticle \cite{AL82} and in the Nambu--Goto string and
Green--Schwarz superstring models, while it remains the symmetry
of the massless particle and the Brink--Schwarz superparticle, as
well as of the tensionless branes and superbranes \cite{null}.

The Hamiltonian analysis of the supertwistor formulation is
performed in Secs.~\ref{twisthamilton} and \ref{unnormalized}. The
generalization of the model to the super--$p$--brane case is given
in Sec.~\ref{pbrane}, and conclusions are given in
Sec.~\ref{conclusions}.

\section{A new supersymmetric string action in
the maximally enlarged superspace } \label{action}

A supersymmetric string in $\Sigma^{({n(n+1)\over 2}|n)}$ is described by
worldsheet functions $X^{\alpha\beta}(\xi)$, $\theta^\alpha(\xi)$,
where $\xi=(\tau,\sigma)$ are the worldsheet $W^2$ coordinates. We
propose the following action:
\begin{eqnarray}\label{St}
S &= & {1\over \alpha^\prime} \int_{W^2} [e^{++} \wedge
\Pi^{\alpha\beta}\, \lambda^-_{\alpha}\lambda^-_{\beta} \nonumber
\\ & &- e^{--} \wedge \Pi^{\alpha\beta}\,
\lambda^+_{\alpha}\lambda^+_{\beta} -
 e^{++} \wedge e^{--}] \; ,
\end{eqnarray}
where
\begin{eqnarray}\label{Pi} &
\Pi^{\alpha\beta}(\xi) = dX^{\alpha\beta}(\xi) - i
d\theta^{(\alpha}\,\theta^{\beta )}(\xi) =
\\ \nonumber
& = d\tau \Pi_{\tau}^{\alpha\beta} + d\sigma
\Pi_{\sigma}^{\alpha\beta} \; ;
\\ \nonumber
& \alpha, \beta =1,\ldots, n\; , \qquad m=0,1\, , \quad
\xi^m=(\tau, \sigma)\;  ,
\end{eqnarray}
and $[1/\alpha^\prime]=\mathrm{ML}^{-1}\,$,
$[\Pi^{\alpha\beta}]=\mathrm{L}\,$, $[e^{\pm\pm}]=\mathrm{L}\;
(c=1)$. The two auxiliary worldvolume fields, the {\it bosonic}
spinors $\lambda^-_{\alpha}(\xi), \lambda^+_{\alpha}(\xi)$, are
{\it dimensionless} and constrained by
\begin{eqnarray}\label{l+l-}
C^{\alpha\beta} \lambda^+_{\alpha}\lambda^-_{\beta} = 1 \; ;
\end{eqnarray}
$e^{\pm\pm}(\xi) = d\xi^m e^{\pm\pm}_m(\xi) = d\tau
e^{\pm\pm}_\tau(\xi) + d\sigma e^{\pm\pm}_\sigma(\xi)$  are two
auxiliary worldvolume one--forms. The one--forms $e^{\pm\pm}$ are
assumed to be linearly independent and, hence, define an auxiliary
worldsheet zweibein
\begin{eqnarray}\label{vielbein}
e^a &=& (e^0, e^1) = d\xi^m e_m^a(\xi) \nonumber \\
&=& ({1\over 2} (e^{++}+ e^{--}),  {1\over 2} (e^{++}- e^{--})).
\end{eqnarray}
The $C^{\alpha\beta}$ in (\ref{l+l-}) is an invertible constant
antisymmetric  matrix
\begin{eqnarray}\label{Cab}
 C^{\alpha\beta}= -  C^{\beta\alpha}\; , \qquad d C^{\alpha\beta}=0\;
 ,
\end{eqnarray}
which can be used to rise and lower the spinor indices (as the
charge conjugation matrix in Minkowski spacetimes). The
invertibility of the matrix $C^{\alpha\beta}$ requires $n$ to be
even; this is not really a limitation since, after all, we are
interested in $n=2^l$ to allow for a spinor treatment of the
$\alpha, \beta$ indices.

For $n=32$ the presence of $C^{\alpha\beta}$ hampers a possible
$D=10$, type IIB treatment of our model. This would require a
$C^{\check{\alpha} I\,  \check{\beta}J}= - C^{\check{\beta}J\,
\check{\alpha}I}$ constructed from the $16\times 16$
Majorana--Weyl sigma matrices and a $2\times2$ matrix in a Lorentz
covariant manner, and there is not a $16\times 16$ charge
conjugation matrix in the $D=10$ Majorana--Weyl representation. As
a result we shall refer to our $n=32$ model
as a  
{\it supersymmetric string in the enlarged $D=11$ superspace 
$\Sigma^{(528|32)}$}, which implies the decomposition of
Eq.~(\ref{n32}). Nevertheless, the $n=32$ case also admits a
$D=10$, type IIA treatment, which uses the same $C^{\alpha\beta}$
of the $D=11$ case, and in which the decomposition (\ref{n32}) is
replaced by its $D=10$, IIA counterpart obtained from (\ref{n32})
by separating the eleventh value of the vector index.

The action (\ref{St}) is invariant under the supersymmetry
transformations
\begin{eqnarray} \label{susy}
& \delta_\epsilon
X^{\alpha\beta}=i\theta^{(\alpha}\epsilon^{\beta)} \quad , \quad
\delta_\epsilon\theta^\alpha=\epsilon^\alpha \quad , \\
\label{susysinglets} & \delta_\epsilon \lambda^\pm_\alpha=0 \quad
, \quad \delta_\epsilon e^{\pm\pm}=0 \quad ,
\end{eqnarray}
as well as under rigid $Sp(n)$ `rotations' acting on the $\alpha,
\beta$ indices.

Note also that, although formally the action (\ref{St}) possesses
a manifest $GL(n,\mathbb{R})$ invariance, the constraint
(\ref{l+l-}) breaks it down to $Sp(n)\subset GL(n,\mathbb{R})$.
Under the action of $Sp(n)$, the Grassmann coordinate functions
$\theta^\alpha(\xi)$ and the auxiliary fields
$\lambda^{\pm}_\alpha(\xi)$ are transformed as symplectic vectors
and $X^{\alpha\beta}(\xi)$ as a symmetric symplectic tensor.
Nevertheless, we keep for them the `spinor' and `spin--tensor'
terminology having in mind their transformation properties under
the subgroup $Spin(t,D-t) \subset Sp(n)$, which would appear in a
`standard' $(t,D-t)$ spacetime treatment.

The above $\Sigma^{({n(n+1)\over 2}|n)}$ supersymmetric string model may
also be described by an action written in terms of {\it
dimensionful} unconstrained spinors $\Lambda^\pm_\alpha(\xi)$,
$[\Lambda^\pm_\alpha]=(\mathrm{ML}^{-1})^{1/2}$,
\begin{eqnarray}\label{St01}
S & = \int_{W^2} [e^{++} \wedge \Pi^{\alpha\beta}\,
\Lambda^-_{\alpha}\Lambda^-_{\beta} - e^{--} \wedge
\Pi^{\alpha\beta}\, \Lambda^+_{\alpha}\Lambda^+_{\beta} \nonumber
\\ &  - \alpha^\prime e^{++} \wedge e^{--}
(C^{\alpha\beta}\Lambda^+_{\alpha}\Lambda^-_{\beta})^2] \; .
\end{eqnarray}
Indeed, one can see that the action (\ref{St01}) possesses two
independent scaling gauge symmetries defined by the transformation
rules
\begin{eqnarray}\label{sc+}
e^{++}(\xi)\rightarrow e^{2\alpha (\xi)} e^{++}(\xi)\; , \quad
\Lambda^-_{\alpha}(\xi) \rightarrow e^{-\alpha (\xi)}
\Lambda^-_{\alpha}(\xi)
\end{eqnarray}
and
\begin{eqnarray}\label{sc-}
e^{--}(\xi)\rightarrow e^{2\beta (\xi)} e^{--}(\xi)\; , \quad
\Lambda^+_{\alpha}(\xi) \rightarrow e^{-\beta (\xi)}
\Lambda^+_{\alpha}(\xi)\; .
\end{eqnarray}
This allows one to obtain
$C^{\alpha\beta}\Lambda^+_{\alpha}\Lambda^-_{\beta}={1 /
\alpha^\prime }$ as a gauge fixing condition. Then the gauge fixed
version of the action (\ref{St01}) coincides with (\ref{St}) up to
the trivial redefinition $\Lambda^{\pm}_{\alpha}=
(\alpha^\prime)^{-1/2}\lambda^{\pm}_{\alpha}$. The gauge
$C^{\alpha\beta}\Lambda^+_{\alpha}\Lambda^-_{\beta}={1 /
\alpha^\prime }$ (equivalent to Eq.~(\ref{l+l-})) is preserved by
a one-parametric combination of (\ref{sc+}) and (\ref{sc-}) with
$\alpha=- \beta$, which is exactly the $SO(1,1)$ gauge symmetry
(worldvolume Lorentz symmetry) of the action (\ref{St}),
\begin{eqnarray}\label{SO(1,1)}
e^{\pm\pm}(\xi)\rightarrow e^{\pm 2\alpha (\xi)} e^{\pm\pm}(\xi)\;
, \;\; \lambda^{\pm}_{\alpha}(\xi) \rightarrow e^{\pm\alpha (\xi)}
\lambda^{\pm}_{\alpha}(\xi)\,.
\end{eqnarray}

The tension parameter $T=1/\alpha^\prime$ enters in the last
(`cosmological') term of the action (\ref{St01}) only. Setting in
it $\alpha^\prime=0$ one finds that the model is non-trivial only
for $e^{++} \propto e^{--}$ and $\Lambda^+ \propto \Lambda^-$ in
which case one arrives at the tensionless super--$p$--brane action
of ref. \cite{B02}, $S=\int d^2\xi \rho^{++m} \Pi^{\alpha\beta}_m
\Lambda^-_\alpha \Lambda^-_\beta$. As we are not interested in
this case, we set $\alpha^\prime=1$ below since the
$\alpha^\prime$ factors can be easily restored by dimensional
considerations.

The most interesting feature of the model (\ref{St}), (\ref{St01})
is that, being formulated in the maximal $\Sigma^{({n(n+1)\over
2}|n)}$ superspace with $n$ fermionic coordinates, it possesses
$(n-2)$ $\kappa$--symmetries; we will prove this in the next
section. For a supersymmetric extended object in standard
superspace, the $\kappa$--symmetry of its worldvolume action
determines the number $k$ of supersymmetries which are preserved
by the ground state (which is a $\nu=\frac{k}{n}$ BPS state made
out of $\#_p=n-k$ preons if at least one supersymmetry, $k \ge 1$,
is preserved). In the present case, we may expect that the ground
state of our model should preserve ($n-2$) out of $n$
supersymmetries, {\it i.e.} is a ${n-2\over n}$ BPS state
($\#_p=2, {30\over 32}$ BPS state for the $D=11$ maximal
superspace $\Sigma^{(528|32)}$).

For $n=2$, $X^{\alpha\beta}$ provides a representation of the
3--dimensional Minkowski space coordinates, $X^{\alpha\beta}
\propto \gamma_\mu^{\alpha\beta}x^\mu$ ($\alpha, \beta = 1,2 $;
$\mu=0,1,2$). Thus the $n=2$ model (\ref{St}) describes a string
in the $D=3$ standard $\Sigma^{(3|2)}$ superspace. However, in the
light of the above discussion, it does not possess any
$\kappa$--symmetry and, hence, its ground state is not a BPS state
since it does not preserve any supersymmetry.

The situation becomes different starting with the $n=4$ model
(\ref{St}), which possesses two $\kappa$--symmetries, the same
number as the Green--Schwarz superstring in the standard $D=4$
superspace. For $D \ge 6$, $n \ge 8$ the number of
$\kappa$-symmetries of our model exceeds $n/2$ and thus the model
describes the excitations of BPS states with
 `supernumerary' supersymmetries  \cite{CLP02}, a ${30\over 32}$
BPS state in the $D=11$ $\Sigma^{(528|32)}$ superspace.

The number of {\it bosonic} degrees of freedom (the number of the
bosonic chiral fields) of our model is $4n-6$ (Sec.~\ref{hm}). It
is not as large as it might look at first sight due to the
`momentum space  dimensional reduction mechanism' \cite{BLS99}
which occurs due to the presence of auxiliary  spinor variables
entering the generalized Cartan--Penrose relation (Eq.~(\ref{PX}))
generated by our model. However, it is larger than that of the
$(D=3,4,6,10)$ Green--Schwarz superstring (which has $D$
[$2n=4(D-2)$] bosonic [fermionic] configuration space real degrees
of freedom, which reduce to $D-2$ [$2(D-2)$] after taking into
account reparametrization invariance ($\kappa$--symmetry), thus
resulting in $2(D-2)$ bosonic and $2(D-2)$ fermionic phase space
degrees of freedom). This, in the light of the above mentioned
relation of the models in maximal superspaces with higher spin
theories, allows us to consider our model as a higher spin
generalization of the Green--Schwarz  superstring, containing
additional information about the nonperturbative states of the
String/M-theory.

The number of {\it fermionic} degrees of freedom of our model is
$2$ for any $n$, less than that of the $D=4,6,10$ $(N=2)$
Green--Schwarz superstring.

\section{Properties of the 
 $\Sigma^{({n(n+1)\over 2}|n)}$ 
supersymmetric string model}
\label{properties}

\subsection{Equations of motion}

Consider the variation of the action (\ref{St}). Allowing for
integration by parts one finds
\begin{eqnarray}\label{variation}
\delta S =  & \int_{W^2} d( e^{--}
 \lambda^+_{\alpha}\lambda^+_{\beta}-
e^{++}\lambda^-_{\alpha}\lambda^-_{\beta} ) \,  i_\delta
 \Pi^{\alpha\beta} \nonumber \\
& -  2i \int_{W^2} e^{++} \wedge d \theta^\alpha
\lambda^-_\alpha \; \delta\theta^\beta \lambda^-_\beta \nonumber\\
& +  2i \int_{W^2} e^{--} \wedge d \theta^\alpha
\lambda^+_\alpha \; \delta\theta^\beta \lambda^+_\beta \nonumber\\
& +  \int_{W^2} (\Pi^{\alpha\beta}\,
\lambda^+_{\alpha}\lambda^+_{\beta} -e^{++}) \wedge \delta
e^{--}\nonumber
\\
& -  \int_{W^2} (\Pi^{\alpha\beta}
\lambda^-_{\alpha}\lambda^-_{\beta} - e^{--}) \wedge \delta e^{++}
 \nonumber
\\
& +    \delta_{\lambda} S \; ,
\end{eqnarray}
where $i_\delta \Pi^{\alpha\beta} \equiv \delta X^{\alpha\beta} -
i \delta \theta^{(\alpha}\theta^{\beta)}$ and the last term
\begin{eqnarray}\label{varl}
 \delta_{\lambda} S =& + \int_{W^2}
2e^{++}\wedge\Pi^{\alpha\beta}
\lambda^-_{\beta} \delta \lambda^-_{\alpha}\nonumber \\
& - \int_{W^2} 2e^{--}\wedge\Pi^{\alpha\beta} \lambda^+_{\beta}
\delta \lambda^+_{\alpha} \; ,
\end{eqnarray}
collects the variations of the bosonic spinors
$\lambda^\pm_{\alpha}(\xi)$.

One easily finds that the equations of motion for the bosonic
coordinate functions, $\delta S / \delta X^{\alpha\beta} (= \delta
S / i_\delta \Pi^{\alpha\beta})=0$ restrict the auxiliary spinors
and auxiliary one--forms,
\begin{eqnarray}\label{vX}
d( e^{--}
 \lambda^+_{\alpha}\lambda^+_{\beta}-
e^{++}\lambda^-_{\alpha}\lambda^-_{\beta} ) =0  \, . \qquad
\end{eqnarray}
The equations for the fermionic coordinate functions, $\delta S /
\delta \theta^{\alpha}=0$, read
\begin{eqnarray}
\label{vTh0} & e^{++} \wedge d \theta^\alpha \lambda^-_\alpha
\lambda^-_\beta - e^{--} \wedge d \theta^\alpha \lambda^+_\alpha
\lambda^+_\beta =0 \; ,
\end{eqnarray}
which, due to the linear independence of the spinors
$\lambda^+_{\alpha}$ and $\lambda^-_{\alpha}$, imply
\begin{eqnarray}
\label{vTh} e^{++} \wedge d \theta^\alpha \lambda^-_\alpha =0\; ,
\qquad e^{--} \wedge d \theta^\alpha \lambda^+_\alpha =0 \; .
\end{eqnarray}
The equations for the one--forms $e^{\pm\pm}(\xi)$ express them
through the worldsheet covariant bosonic form (\ref{Pi}) of the
$\Sigma^{({n(n+1)\over 2}|n)}$ superspace and the spinors
$\lambda^\pm_{\alpha}(\xi)$,
\begin{eqnarray}
\label{ve-} e^{++} & = & \Pi^{\alpha\beta}\,
\lambda^+_{\alpha}\lambda^+_{\beta} \; , \qquad
\\ \label{ve+}
e^{--} & = & \Pi^{\alpha\beta} \lambda^-_{\alpha}\lambda^-_{\beta}
\; . \qquad
\end{eqnarray}
This reflects the auxiliary nature of $e^{\pm\pm}$ and implies
that Eqs.~(\ref{vX}) and (\ref{vTh}) actually restrict
$\Pi^{\alpha\beta}$ and $d\theta^\alpha$,
\begin{eqnarray}\label{vXs}
d( \Pi^{\gamma\delta} \lambda^-_{\gamma}\lambda^-_{\delta} \;
\lambda^+_{\alpha}\lambda^+_{\beta}- \Pi^{\gamma\delta}
\lambda^+_{\gamma}\lambda^+_{\delta} \;
\lambda^-_{\alpha}\lambda^-_{\beta} ) & = & 0  \, , \qquad
\\ \label{vThs-}
\Pi^{\gamma\delta} \lambda^+_{\gamma}\lambda^+_{\delta} \wedge d
\theta^\alpha \lambda^-_\alpha & = & 0\; , \qquad
\\ \label{vThs+}
 \Pi^{\gamma\delta} \lambda^-_{\gamma}\lambda^-_{\delta}
\wedge d \theta^\alpha \lambda^+_\alpha & = & 0 \; . \; \qquad
\end{eqnarray}

Moreover, looking at Eqs.~(\ref{ve-}), (\ref{ve+}) one can easily
see the necessity of the constraints (\ref{l+l-}) on the bosonic
spinor variables. Indeed, if one would ignore these constraints
and vary the action with respect to unconstrained
$\lambda^\pm_\alpha$, one would arrive, from (\ref{varl}), at
$e^{++}\wedge \Pi^{\alpha\beta}\, \lambda^-_{\beta}=0$ and
$e^{--}\wedge \Pi^{\alpha\beta}\, \lambda^+_{\beta}=0$. By
(\ref{ve-}) (or (\ref{ve+})) this would imply, in particular,
$e^{++}\wedge e^{--}=0$, contradicting
 the original assumption of independence of the one--forms $e^{++}$ and
$e^{--}$ and, actually, reducing the present model to a $p=1$
version of the tensionless $p$--brane model \cite{B02}.

As  $\lambda^{\pm}_\alpha$ are restricted by the constraint
(\ref{l+l-}), this constraint has to be taken into account in the
variational problem. Instead of applying the Lagrange multiplier
technique, one may restrict the variations to those that preserve
(\ref{l+l-}), {\it i.e.} such that
\begin{equation}\label{vl+l-}
C^{\alpha\beta} \delta \lambda^+_{\alpha}\lambda^-_{\beta} +
C^{\alpha\beta} \lambda^+_{\alpha} \delta\lambda^-_{\beta}=0\; .
\end{equation}
One can solve (\ref{vl+l-}) by introducing a set of $n-2$
auxiliary spinors $u^I_{\alpha}$ `orthogonal' to the
$\lambda^{\pm}$ ({\it cf.}~\cite{BZ95,BZ}),
\begin{equation} \label{cond1}
C^{\alpha\beta} u_\alpha^I \lambda^{\pm}_\beta = 0 \; , \qquad
I=1, \ldots, n-2 \quad ,
\end{equation}
and normalized by
\begin{equation} \label{cond2}
C^{\alpha\beta} u_\alpha^I u_\beta^J = C^{IJ} \, , \qquad
C^{IJ}=-C^{JI} \; ,
\end{equation}
where $C^{IJ}$ is an antisymmetric constant invertible $(n-2)
\times (n-2)$ matrix.

The $n$ spinors
\begin{equation} \label{basis}
\{\lambda^+_\alpha\; , \, \lambda^-_\alpha \; , \quad u^I_\alpha\}
\; , \; I=1, \ldots, n-2 \; ,
\end{equation}
provide a basis that can be used to decompose an arbitrary spinor
worldvolume function ({\it cf.} \cite{NP85}), and in particular
the variations $\delta \lambda^+$, $\delta\lambda^-$. Then one
finds that the only consequence of Eq.~(\ref{vl+l-}) is that the
sum of the coefficient for $\lambda^+$ in the decomposition of
$\delta \lambda^+$  and that of $\lambda^-$ in the decomposition
of $\delta \lambda^-$ vanishes . In other words, the general
solution of Eq.~(\ref{vl+l-}) reads
\begin{eqnarray}\label{vlpm}
\delta \lambda_\alpha^+ &=& \omega(\delta) \lambda_\alpha^+ +
\Omega^{++}(\delta)\lambda_\alpha^- +
\Omega^+_I(\delta)u_\alpha^I \; , \\
\label{vlpm2} \delta \lambda_\alpha^- &=& - \omega(\delta)
\lambda_\alpha^- + \Omega^{--}(\delta)\lambda_\alpha^+ +
\Omega^-_I(\delta)u_\alpha^I \; ,
\end{eqnarray}
where $\Omega^{\pm}_I(\delta)$, $\Omega^{\pm\pm}(\delta)$ and
$\omega(\delta)$ are arbitrary variational parameters.

Substituting Eqs.~(\ref{vlpm}), (\ref{vlpm2}) into (\ref{varl}),
one finds {\setlength\arraycolsep{2pt}
\begin{eqnarray}\label{var2}
 \delta_{\lambda} S = & \!\!\!\! -  \int_{W^2}
(2e^{++}\wedge\Pi^{\alpha\beta} \lambda^-_{\beta}
\lambda^-_{\alpha}  \qquad
\nonumber \\
& +  2e^{--}\wedge\Pi^{\alpha\beta} \lambda^+_{\beta}
\lambda^+_{\alpha}) \omega(\delta)  \nonumber \\
& +  \int_{W^2} 2e^{++}\wedge\Pi^{\alpha\beta} \lambda^-_{\beta}
\lambda^+_{\alpha} \Omega^{--}(\delta)
\nonumber \\
& + \int_{W^2} 2e^{--}\wedge\Pi^{\alpha\beta} \lambda^+_{\beta}
\lambda^-_{\alpha} \Omega^{++}(\delta)
\nonumber \\
& +  \int_{W^2} 2e^{++}\wedge\Pi^{\alpha\beta} \lambda^-_{\beta}
u_{\alpha}^I \Omega^-_I(\delta)
\nonumber \\
& -   \int_{W^2} 2e^{--}\wedge\Pi^{\alpha\beta} \lambda^+_{\beta}
u^I_{\alpha} \Omega^+_I(\delta)\; .
\end{eqnarray}}

Now we can easily write the complete set of equations of motion
which include, in addition to Eqs.~(\ref{vX}), (\ref{vTh}),
(\ref{ve-}), (\ref{ve+}), the set of equations for
$\lambda^{\pm}_\alpha$,  which follows from $\delta S /
\omega(\delta)=0$, $\delta S / \Omega^{++}(\delta)=0$, $\delta S /
\Omega^+_I(\delta)=0$, $\delta S / \Omega^{--}(\delta)=0$, and
$\delta S / \Omega^-_I(\delta)=0$, namely
\begin{eqnarray}
\label{vl0} & e^{++}\wedge \Pi^{\alpha\beta} \lambda^-_{\beta}
\lambda^-_{\alpha} + e^{--}\wedge\Pi^{\alpha\beta}
\lambda^+_{\beta}\lambda^+_{\alpha} =0 \; ,
\\ \label{vl--}
& e^{++}\wedge\Pi^{\alpha\beta} \lambda^-_{\beta}
\lambda^+_{\alpha} =0 \; ,
\\ \label{vl++}
& e^{--}\wedge\Pi^{\alpha\beta} \lambda^+_{\beta}
\lambda^-_{\alpha} =0 \; ,
\\ \label{vl-I}
& e^{++}\wedge\Pi^{\alpha\beta} \lambda^-_{\beta} u_{\alpha}^I =0
\; ,
\\ \label{vl+I}
& e^{--}\wedge\Pi^{\alpha\beta} \lambda^+_{\beta} u^I_{\alpha} =0
\; .
\end{eqnarray}
Due to the linear independence of $e^{++}=d\xi^m e_m^{++}(\xi)$
and $e^{--}=d\xi^m e_m^{--}(\xi)$, Eqs.~(\ref{vl--}), (\ref{vl++})
imply
\begin{eqnarray}\label{P+-=0}
\Pi^{\alpha\beta} \lambda^-_{\beta}
\lambda^+_{\alpha} =0 \; .
\end{eqnarray}
Decomposing the bosonic invariant one form $\Pi^{\alpha\beta}=
d\xi^m \Pi^{\alpha\beta}_m$ in the (`unholonomic') basis provided
by $e^{\pm\pm}$,
 \begin{eqnarray}\label{P=eP}
\Pi^{\alpha\beta}
= e^{++} \Pi^{\alpha\beta}_{++} + e^{--}  \Pi^{\alpha\beta}_{--} \; ,
\\ \label{Pe++}
\Pi^{\alpha\beta}_{\pm\pm}=\nabla_{\pm\pm} X^{\alpha\beta} - i
\nabla_{\pm\pm} \theta^{(\alpha } \; \theta^{\beta ) } \; ,
\end{eqnarray}
where $\nabla_{\pm\pm}$ is defined by
\begin{equation} \label{d=en} d \equiv e^{\pm\pm} \nabla_{\pm\pm}
= e^{++} \nabla_{++} + e^{--} \nabla_{--}\; ,
\end{equation}
 one finds that Eqs.~(\ref{vl-I}) and (\ref{vl+I}) restrict
only the left and right chiral derivatives
$(\nabla_{++},\,\nabla_{--})$ of the bosonic coordinate function
$X^{\alpha\beta}(\xi)$, respectively,
\begin{eqnarray}
 \label{vl-I--}
\Pi_{--}^{\alpha\beta} \lambda^-_{\beta} u_{\alpha}^I
&\equiv&
(\nabla_{--} X^{\alpha\beta} -
i  \nabla_{--} \theta^{(\alpha } \; \theta^{\beta ) })
\; \lambda^-_{\beta} u_{\alpha}^I
=0 \quad , \nonumber \\ {}
\\ \label{vl+I++}
\Pi_{++}^{\alpha\beta} \lambda^+_{\beta} u^I_{\alpha}
&\equiv& (\nabla_{++} X^{\alpha\beta} -
i  \nabla_{++} \theta^{(\alpha } \; \theta^{\beta ) } )
\; \lambda^+_{\beta} u_{\alpha}^I
=0 \quad .  \nonumber \\ {}
\end{eqnarray}
In the same manner, Eqs.~(\ref{vTh}) can be written as
\begin{eqnarray}
\label{vTh--}
\nabla_{--}
\theta^\alpha \, \lambda^-_\alpha =0\; ,
\qquad \nabla_{++}\theta^\alpha \, \lambda^+_\alpha =0 \; .
\end{eqnarray}

\noindent The analysis of the above set of equations in the
maximal superspace, the search for solutions and their
reinterpretation in standard $D$--dimensional spacetime (possibly,
along the fields--extended superspace democracy of \cite{JdA00},
or of the `two--time physics' \cite{Bars}) is a problem for future
study.

\subsection{Gauge symmetries}

The expression (\ref{variation}), with (\ref{var2}), for the
general variation of the $\Sigma^{(n(n+1)|n)}$ 
supersymmetric string action (\ref{St})
shows that the model possesses $n$ supersymmetries and $(n-2)$
$\kappa$--symmetries of the form
\begin{eqnarray}
\label{kappa1} & \delta_\kappa \theta^\alpha(\xi) =
C^{\alpha\beta}u^I_\beta (\xi) \kappa_I(\xi) \; ,\\
\label{kappa2} & \delta_\kappa X^{\alpha\beta}(\xi)
= i  \delta_\kappa \theta^{(\alpha}(\xi)\theta^{\beta)}(\xi) \; ,\\
\label{kappa3} & \delta_\kappa \lambda_\alpha^\pm(\xi)=0\; ,
\qquad
 \delta_\kappa e^{\pm\pm}_m(\xi)=0 \; ,
\end{eqnarray}
with $(n-2)$ fermionic gauge parameters $\kappa_I(\xi)$ (30 for
$\Sigma^{(528|32)})$. In the framework of the second Noether
theorem this $\kappa$--symmetry is reflected by the fact that only
2 of the $n$ fermionic equations (\ref{vTh0}) are independent. We
stress that the ($n-2$) $GL(n,\mathbb{R})$ vector fields
$u^I_\alpha$ defined by (\ref{cond1}) are auxiliary. They allow us
to write explicitly the general solution of the equations
\begin{equation}\label{kappa4}
\delta_\kappa \theta^\alpha(\xi)\lambda^{\pm}_\alpha(\xi)=0 \quad
,
\end{equation}
which define implicitly the  $\kappa$--symmetry transformation
(\ref{kappa1}). Note that the dynamical system is
$\kappa$--symmetric despite it does not contain a Wess--Zumino
term. This property seems to be  specific of models defined on
maximal superspaces.

Our model also possesses ${1 \over 2}(n-1)(n-2)$ $b$-symmetries,
which are the bosonic `superpartners' of the fermionic
$\kappa$--symmetries, defined by
\begin{eqnarray}
& \delta_b X^{\alpha\beta}=b_{IJ}(\xi)u^{\alpha I}u^{\beta J} \; ,
\nonumber \\ \label{b4} & \delta_b \theta^\alpha= 0\; ,\qquad
\delta_b \lambda^\pm_\alpha=0 \;,\qquad \delta_b e^{\pm\pm}=0 \; ,
\end{eqnarray}
where $b_{IJ}(\xi)$ is symmetric and $I,J=1,\ldots,n-2$. They are
reflected by  the $(n-1)(n-2)/2$ Noether identities stating that
the contractions of the bosonic equations (\ref{vX}) with the
$u^{\alpha I}u^{\beta J}$ bilinears of the $(n-2)$ auxiliary
bosonic spinors $u^{\alpha I}(=C^{\alpha\beta}u^I_\beta)$ vanish
\footnote{In the massless $\Sigma^{(n(n+1)|n)}$ superparticle and
tensionless super--$p$--brane models the $b$--symmetry
\cite{BL98,BLS99}, \cite{B02} ({\it cf.}~\cite{ZU02}) is
$n(n-1)/2$ parametric. This comes from the fact that such models
contain a single bosonic spinor ${\lambda}_\alpha$ and the
nontrivial $b$--symmetry variation is the general solution of the
spinorial equation 
 $\delta_b X^{\alpha\beta} {\lambda}_\alpha=0$.
In our tensionful $\Sigma^{(n(n+1)|n)}$ supersymmetric string 
model with two bosonic spinors
$\lambda^{\pm}_\alpha(\xi)$, the $(n-1)(n-2)/2$ parametric
$b$--symmetry transformations (Eq. (\ref{b4})) are the solutions
of two equations
 $\delta_b X^{\alpha\beta} {\lambda}^+_\alpha=0$ and
 $\delta_b X^{\alpha\beta} {\lambda}^-_\alpha=0$.}.

The remaining gauge symmetries of the action (\ref{St}) are the
$SO(1,1)$ worldsheet Lorentz invariance
\begin{eqnarray}\label{sc1}
\delta X^{\alpha\beta}=0\, &,& \, \delta\theta^\alpha =0 \;, \nonumber\\
\delta\lambda_\alpha^\pm=
\pm\omega(\delta)\lambda_\alpha^\pm\;&,&\; \delta e^{\pm\pm}=\pm
2\omega(\delta) e^{\pm\pm} \; ,
\end{eqnarray}
which is reflected by the fact that Eq.~(\ref{vl0}) is satisfied
identically when Eqs.~(\ref{ve-}), (\ref{ve+}) are taken into
account,
 and the symmetry under worldvolume general coordinate transformations.

As customary in  string models, the general coordinate invariance
and the $SO(1,1)$ gauge symmetry allows one to fix locally the
conformal gauge where $e_m{}^a(\xi) = e^{\phi(\xi)} \delta_m^a$
or, equivalently
\begin{eqnarray}\label{cg}
e^{++}=  e^{\phi(\xi)} (d\tau + d\sigma)\; , \qquad e^{--}=
e^{\phi(\xi)} (d\tau - d\sigma)\; , \quad
\\ \label{cgc}
\Leftrightarrow \quad e^{++}_\sigma = e^{++}_\tau = e^{\phi(\xi)}
\; , \quad e^{--}_\sigma = - e^{--}_\tau = - e^{\phi(\xi)} \;
.\quad
\end{eqnarray}
This indicates that it makes sense to consider the fields
$e_\sigma^{\pm\pm}(\tau,\sigma)$ as nonsingular (${1\over
e_\sigma^{\pm\pm}}= \pm e^{-\phi(\xi)}$ in the conformal gauge), a
fact used in the Hamiltonian analysis below.

There is a correspondence \cite{BKOP97,ST97} between the
$\kappa$--symmetry of the worldvolume action and the supersymmetry
preserved by a BPS state ({\it e.g.} by a solitonic solution of
the supergravity equations of motion). Thus, the action (\ref{St})
defines a dynamical model for the excitations of a BPS state
preserving {\it all but two} supersymmetries. Such a BPS state can
be treated as a composite of two BPS preons ($\#_p = 32-30$). This
will become especially transparent after the Hamiltonian analysis
of next section.

\section{Hamiltonian mechanics}\label{hm} \label{Hamiltonian}

The gauge symmetry structure has already been shown in the
Lagrangian framework. However, our dynamical system clearly
possesses additional, second class, constraints \cite{Dirac}, one
of which is condition~(\ref{l+l-}). In this section we carry out
the Hamiltonian analysis of our $\Sigma^{({n(n+1)\over 2}|n)}$
supersymmetric string model. In particular, this will allow us to find the
number of field theoretical degrees of freedom and to establish
the relation of our model with the notion of BPS preons
\cite{BPS01}.

The Lagrangian density ${\cal L}$ for the action (\ref{St}),
\begin{equation}\label{StL}
S  =  \int_{W^2} d\tau d\sigma \; {\cal L}
\; ,
\end{equation}
is given by
\begin{eqnarray}\label{St2}
{\cal L} & = & (e^{++}_{\tau} \Pi_{\sigma}^{\alpha\beta} -
e^{++}_{\sigma} \Pi_{\tau}^{\alpha\beta} )
\lambda^-_{\alpha}\lambda^-_{\beta}  \qquad
\nonumber \\
& - &  (e^{--}_{\tau} \Pi_{\sigma}^{\alpha\beta} - e^{--}_{\sigma}
\Pi_{\tau}^{\alpha\beta} ) \lambda^+_{\alpha}\lambda^+_{\beta}
\nonumber  \\
& - & (e^{++}_{\tau} e^{--}_{\sigma} - e^{++}_{\sigma}
e^{--}_{\tau}) \; ,
\end{eqnarray}
where
\begin{eqnarray}
\Pi_{\tau}^{\alpha\beta} =
\partial_{\tau}X^{\alpha\beta} -i\partial_{\tau}
\theta^{(\alpha} \theta^{\beta)} \, , \qquad \nonumber \\
\label{Pits} \Pi_{\sigma}^{\alpha\beta} =
\partial_{\sigma}X^{\alpha\beta} -i\partial_{\sigma}
\theta^{(\alpha} \theta^{\beta)} \, ,  \qquad
\end{eqnarray}
are the worldsheet components of the one-form (\ref{Pi}).

The momenta $P_{{\cal M}}$ canonically conjugate to the
configuration space variables
\begin{eqnarray}\label{cZcM}
{\cal Z}^{{\cal M}}  \equiv {\cal Z}^{{\cal M}}(\tau, \sigma) :=
\left( X^{\alpha\beta} \, , \, \theta^\alpha  , \lambda^\pm_\alpha
\, , \, e^{\pm\pm}_\tau \, , \, e^{\pm\pm}_\sigma \right)\;
\end{eqnarray}
are defined as usual:
\begin{eqnarray}\label{cPcM}
P_{{\cal M}} =(P_{\alpha\beta}\,,\, \pi_\alpha \, , \, P_{\pm
}^{\alpha (\lambda)} , \, P_{\pm\pm}^{\tau} \,, \,
P_{\pm\pm}^{\sigma}) = {\partial {\cal L} \over \partial
(\partial_\tau {\cal Z}^{{\cal M}})} \; .\quad
\end{eqnarray}

The canonical equal $\tau$ graded Poisson brackets,
$$[{\cal Z}^{{\cal N}}(\sigma) \, , \, P_{{\cal M}} (\sigma^\prime)\}_{_P}
=  - (-1)^{{\cal N}{\cal M}} [ P_{{\cal M}} (\sigma^\prime) \, ,
\, {\cal Z}^{{\cal N}}(\sigma ) \}_{_P} \; , $$ are defined by
\begin{eqnarray}
\label{canonical} [ {\cal Z}^{{\cal N}}(\sigma^{\prime}) \, , \,
P_{{\cal M}}(\sigma ) \}_{_P} &:=& (-1)^{\cal N} \delta^{{\cal
N}}_{{\cal M}} \delta(\sigma-\sigma^{\prime})\; ,
\end{eqnarray}
where $(-1)^{\cal N} \equiv (-1)^{\mathrm{deg}({\cal N})}$ and the
degree $\mathrm{deg}({\cal N})\equiv \mathrm{deg}({\cal Z}^{\cal
N})$ is $0$ for the bosonic fields, ${\cal Z}^{\cal
N}=X^{\alpha\beta}, \lambda^\pm_\alpha, e_m^{\pm\pm}$ (or for the
`bosonic indices' ${\cal N}=(\alpha\beta), (\alpha\pm ), (\pm\pm),
m$), and $1$ for the fermionic fields ${\cal Z}^{\cal
N}=\theta^\alpha$ (or for the `fermionic indices'  ${\cal
N}=\alpha $ and ${\cal N}= \pm$ that we will meet below in the
supertwistor formulation of the model).

Since the action (\ref{St}) is clearly of first order type, it is
not surprising that the expression of every momentum results in a
primary \cite{Dirac} constraint. Explicitly,
\begin{eqnarray} \label{PX}
{\cal P}_{\alpha\beta} & = &  P_{\alpha\beta} +
e_{\sigma}^{++} \lambda^-_\alpha \lambda^-_\beta -
e_{\sigma}^{--} \lambda^+_\alpha \lambda^+_\beta \approx 0 \; , \\
\label{D}
{\cal D}_\alpha& = &  \pi_\alpha + i
\theta^\beta P_{\alpha\beta} \approx 0 \; , \\
\label{Plam}
P_{\pm}^{\alpha (\lambda)} & \approx & 0 \; , \\
\label{Psig}
P_{\pm \pm}^{\sigma} & \approx & 0 \; , \\
\label{Ptau}
P_{\pm\pm}^{\tau} & \approx & 0 \; ,
\end{eqnarray}
where only ${\cal D}_\alpha$ is fermionic. Condition (\ref{l+l-}),
\begin{eqnarray}\label{cl+l-}
{\cal N} := C^{\alpha\beta} \lambda^+_{\alpha}\lambda^-_{\beta}
 - 1 \approx 0 \; ,
\end{eqnarray}
imposed on the bosonic spinors from the beginning, is also a
primary constraint and has to be treated on the same footing as
Eqs.~(\ref{PX})-(\ref{Ptau}).

The {\it canonical} Hamiltonian density ${\cal H}_0$,
\begin{equation}\label{H0}
{\cal H}_0  =  \partial_\tau {\cal Z}^{\cal M} \, P_{\cal M} -
{\cal L} \quad ,
\end{equation}
calculated on the primary constraints (\ref{PX})--(\ref{Ptau})
hypersurface reads
\begin{eqnarray}\label{H1}
{\cal H}_0 &=&  e^{--}_{\tau} \Pi^{\alpha\beta}_{\sigma}
\lambda^+_\alpha \lambda^+_\beta -e^{++}_{\tau}
\Pi^{\alpha\beta}_{\sigma} \lambda^-_\alpha \lambda^-_\beta +
\qquad \nonumber
\\
&+& (e^{++}_{\tau} e^{--}_{\sigma} - e^{++}_{\sigma}
e^{--}_{\tau})  \; . \hspace{-1cm}
\end{eqnarray}
The evolution of any functional $f({\cal Z}^{\cal M}, P_{\cal N})$
is defined by $\partial_\tau f = [ f\; , \; \int d\sigma {\cal
H}^\prime ]_{_P}$ involving the total Hamiltonian, $\int d\sigma
{\cal H}^\prime$, where the Hamiltonian density ${\cal H}^\prime$
is the sum of ${\cal H}_0$ in Eq.~(\ref{H1}) and the terms given
by integrals of the primary constraints (\ref{PX})-(\ref{Ptau})
multiplied by arbitrary functions (Lagrange multipliers)
\cite{Dirac}. Then one has to check that the primary constraints
are preserved under the evolution, $\partial_\tau {\cal
P}_{\alpha\beta} \approx 0$, etc. At this stage additional,
secondary constraints may be obtained. This is the case for our
system.

Indeed, since the constraints (\ref{Ptau}) have zero Poisson
brackets with any other primary constraint, their time evolution
is just determined by the canonical Hamiltonian ${\cal H}_0$,
$\partial_\tau{\cal P}_{\pm \pm}^{\tau}= [{\cal P}_{\pm
\pm}^{\tau},\int d \sigma {\cal H}_0 ]_{_P}$. Then one easily sees
that $\partial_\tau{\cal P}_{\pm \pm}^{\tau} \approx 0$ produces a
pair of secondary constraints,
\begin{eqnarray} \label{phipm1}
\Phi_{\pm \pm}&:=& \Pi_{\sigma}^{\alpha \beta} \lambda^\mp_\alpha
\lambda^\mp_\beta - e^{\mp\mp}_\sigma =
\nonumber \\
&=&
(\partial_{\sigma} X^{\alpha \beta}-i
\partial_{\sigma}\theta^{(\alpha}\theta^{\beta)})\lambda^\mp_\alpha
\lambda^\mp_\beta - e^{\mp\mp}_\sigma \approx 0\; . \quad
\end{eqnarray}
Slightly more complicated calculations with the total ${\cal
H}^\prime$ show that we also have the secondary constraint
\begin{eqnarray} \label{phi0}
\Phi^{(0)}&:=& \Pi_{\sigma}^{\alpha \beta} \lambda^+_\alpha
\lambda^-_\beta = \nonumber \\
&=& (\partial_{\sigma} X^{\alpha \beta}-i
\partial_{\sigma}\theta^{(\alpha}\theta^{\beta)})\lambda^+_\alpha
\lambda^-_\beta \approx 0  \qquad
\end{eqnarray}
(details about its derivation can be found below
Eq.~(\ref{contractions})). The appearance of this secondary
constraint may be understood as well by comparing with the results
of the Lagrangian approach: it is just the $\sigma$ component of
the differential form equation (\ref{P+-=0}).

The secondary constraints (\ref{phipm1}) imply that the canonical
Hamiltonian ${\cal H}_0$, Eq.~(\ref{H0}), vanishes on the surface
of constraints (\ref{phipm1}),
\begin{eqnarray}\label{H0=0}
{\cal H}_0 \approx 0  \, ,
\end{eqnarray}
a characteristic property of theories with general coordinate
invariance. Hence the total Hamiltonian reduces to a linear
combination of the constraints (\ref{PX})--(\ref{Ptau}),
(\ref{phipm1}), (\ref{phi0}),
{\setlength\arraycolsep{0pt}
\begin{eqnarray}\label{H=} {\cal H}
&=& - e_\tau^{++} \Phi_{++} + e_\tau^{--} \Phi_{--} + l^{(0)}
\Phi^{(0)} + L^{\alpha\beta} {\cal P}_{\alpha\beta} + \xi^\alpha
{\cal D}_\alpha \nonumber \\
&+&  l^\pm_\alpha P^{\alpha (\lambda )}_{\pm} +
L^{\pm\pm} P_{\pm\pm}^{\sigma}
+ h^{\pm\pm} P_{\pm\pm}^\tau  +
L^{(n)} {\cal N}
\end{eqnarray}}
where $l^{(0)}$, $ L^{\alpha\beta}$, $\xi^\alpha$, $l^\pm_\alpha$,
$L^{\pm\pm}$, $h^{\pm\pm}$,  $L^{(n)}$ {\it and} $\pm
e_\tau^{\pm\pm}$ are Lagrangian multipliers whose form should be
fixed from the preservation of all the primary and secondary
constraints under $\tau$-evolution.

Note  that the constraints (\ref{Ptau}) are trivially first class,
since their Poisson brackets with all the other constraints,
including (\ref{phipm1}) and (\ref{phi0}), vanish. This allows us
to state  that $e_\tau^{\pm\pm}(\xi)$ are not dynamical fields but
rather Lagrange multipliers (as the time component of
electromagnetic potential $A_0$ in electrodynamics). Nevertheless,
the appearance of these Lagrange multipliers from the $\tau$
components of the zweibein $e_m^{\pm\pm}$ put a `topological'
restriction on a possible gauge fixing; in particular the gauge
$e_\tau^{\pm\pm}=0$ is not allowed. Indeed, the nondegeneracy of
the zweibein, assumed from the beginning, reads
\begin{eqnarray}\label{det(e)}
\mathrm{det}\,(e_m^a(z)) \equiv {1\over 2}
(e_\tau^{--}e_\sigma^{++}- e_\tau^{++} e_\sigma^{--} )\not= 0 \; .
\end{eqnarray}
Just due to this restriction, studying the $\tau$-preservation of
the primary constraints, one finds the secondary constraint
(\ref{phi0}).

If by checking the (primary and secondary) constraints
preservation under $\tau$-evolution one finds that some lagrangian
multipliers remain unfixed, then they correspond to {\it first
class constraints} \cite{Dirac} which generate gauge symmetries of
the system through the Poisson brackets. In other words, since the
canonical Hamiltonian vanishes in the weak sense, the total
Hamiltonian is a linear combination of all first class constraints
\cite{Dirac}.

If some of the equations  resulting from the $\tau$-evolution of
the constraints (or their linear combinations) do not restrict the
Lagrangian multiplier, but imply the vanishing of a combination of
the canonical variables, they correspond to new secondary
constraints, which have to be added with new Lagrange multipliers
to obtain a new total Hamiltonian. In this case the check that all
the constraints are preserved under $\tau$--evolution has to be
repeated.

This does not happen for our dynamical system: a further check of
the constraints $\tau$--preservation does not result in the
appearance of new constraints. Indeed, it leads to the following
set of equations for the Lagrange multipliers
\begin{eqnarray}
\label{dtP}
\partial_\sigma ( e_\tau^{--}\lambda^+_{\alpha}  \lambda^+_{\beta}
- e_\tau^{++} \lambda^-_{\alpha}  \lambda^-_{\beta}
+ l^{(0)} \lambda^+_{(\alpha} \lambda^-_{\beta )}) &-&
\nonumber \\
 -2 e_\sigma^{--} \lambda^+_{(\alpha} l^+_{\beta)}
+2 e_\sigma^{++} \lambda^-_{(\alpha} l^-_{\beta)} &+&
\nonumber \\ +
L^{++} \lambda^-_{\alpha} \lambda^-_{\beta} -
L^{--} \lambda^+_{\alpha} \lambda^+_{\beta} &\approx & 0 \; , \qquad
\\ {} \nonumber && \\
\label{dtD} \lambda^-_{\alpha} \, [2i  e^{++}_{\tau}
(\partial_\sigma  \theta \lambda^-) - i l^{(0)} (\partial_\sigma
\theta \lambda^+) &+& \nonumber  \\ +  2i e^{++}_{\sigma}
(\xi\lambda^-)]\, &-&
\quad \nonumber  \\
 - \lambda^+_{\alpha} [2i  e^{--}_{\tau} (\partial_\sigma  \theta
 \lambda^+) + i l^{(0)} (\partial_\sigma  \theta \lambda^-) &+&
\quad \nonumber  \\
 + 2i e^{--}_{\sigma} (\xi\lambda^+)]\,
 &\approx &  0 \; , \quad
\\ {} && \nonumber \\
\label{dtPl+}
-2 e^{--}_{\tau} \Pi^{\alpha\beta}_\sigma \lambda^+_{\beta} -
l^{(0)}\Pi^{\alpha\beta}_\sigma \lambda^-_{\beta}
+ \hspace{1cm} && \nonumber \\   +
2 e^{--}_{\sigma}L^{\alpha\beta}\lambda^+_{\beta} -
 L^{(n)} C^{\alpha\beta}\lambda^-_{\beta}
&\approx&  0 \; , \quad
\\ \nonumber && \\
\label{dtPl-}
2 e^{++}_{\tau} \Pi^{\alpha\beta}_\sigma \lambda^-_{\beta} -
l^{(0)}\Pi^{\alpha\beta}_\sigma \lambda^+_{\beta} - \hspace{1cm}
&& \nonumber \\  - 2
e^{++}_{\sigma}L^{\alpha\beta}\lambda^-_{\beta}
- L^{(n)} C^{\alpha\beta}\lambda^+_{\beta}
&\approx&  0 \; , \quad
\\ {} \nonumber &&
\\
\label{dtPs++}
e^{--}_{\tau} - L^{\alpha\beta}\lambda^-_{\alpha}\lambda^-_{\beta}
\approx 0 \; , \hspace{1cm} &&
\\
\label{dtPs--}
e^{++}_{\tau} - L^{\alpha\beta}\lambda^+_{\alpha}\lambda^+_{\beta}
\approx  0 \; , \hspace{1cm} &&
\\ \label{dtCll}
l^+_{\alpha} C^{\alpha\beta}\lambda^-_{\beta} -
l^-_{\alpha} C^{\alpha\beta}\lambda^+_{\beta}
\approx 0 \; , \hspace{1cm} &&
\end{eqnarray}
\begin{eqnarray}\label{dtP++}
\partial_\sigma L^{\alpha\beta}  \lambda^-_{\alpha}  \lambda^-_{\beta}
+ 2i (\xi\lambda^-) (\partial_\sigma  \theta \lambda^-)
&+& \quad \nonumber \\
+
2 l^-\Pi_\sigma \lambda^- - L^{--}  &\approx&  0 \; , \quad
\\
\label{dtP--}
\partial_\sigma L^{\alpha\beta}  \lambda^+_{\alpha}  \lambda^+_{\beta}
+ 2i (\xi\lambda^+) (\partial_\sigma  \theta \lambda^+)
&+&  \quad \nonumber \\ +
2 l^+\Pi_\sigma \lambda^+
- L^{++}
&\approx&  0 \; , \quad
\\
\label{dtP0}
\partial_\sigma L^{\alpha\beta}  \lambda^+_{\alpha}  \lambda^-_{\beta}
+ i (\xi\lambda^+) (\partial_\sigma  \theta \lambda^-) - \qquad {} &&
\nonumber \\ - i (\xi\lambda^-) (\partial_\sigma  \theta \lambda^+) +
l^+\Pi_\sigma \lambda^- + l^-\Pi_\sigma \lambda^+
&\approx&  0 \; , \quad
\end{eqnarray}
where the weak equality sign is used to stress that one may use
the constraints in solving the above system of equations. For
brevity, in Eqs.~(\ref{dtP})--(\ref{dtP0}) and below we often omit
spinor indices in the contractions
\begin{eqnarray}\label{contractions}
&& (\partial_\sigma  \theta \lambda^\pm)\equiv
\partial_\sigma  \theta^{\beta} \, \lambda^\pm_{\beta}\; , \qquad
(\xi \lambda^\pm)\equiv  \xi^{\beta} \, \lambda^\pm_{\beta}\; , \qquad
\nonumber \\
&& l^\pm\Pi_\sigma \lambda^\pm\equiv
l^\pm_\alpha \Pi^{\alpha\beta}_\sigma \lambda^\pm_{\beta}\; , \qquad
 l^\pm L \lambda^\pm\equiv
l^\pm_\alpha L^{\alpha\beta}_\sigma \lambda^\pm_{\beta}\; . \qquad
\end{eqnarray}

Note that Eqs.~(\ref{dtP})--(\ref{dtCll}) come from the
requirement of $\tau$--preservation of the primary constraints,
while that for the secondary constraints leads to
Eqs.~(\ref{dtP++})--(\ref{dtP0}). Thus the above statement about
the appearance of the secondary constraint (\ref{phi0}) can be
checked by studying Eqs.~(\ref{dtP})--(\ref{dtCll}) with
$l^{(0)}=0$. In this case the contraction of Eq.~(\ref{dtPl+})
with ($-\lambda_\alpha^-$) and of Eq.~(\ref{dtPl-}) with
$\lambda_\alpha^+$ results, respectively, in the equations
$e_\tau^{--} \lambda^+\Pi_\sigma \lambda^- - e_\sigma^{--}
\lambda^+ L \lambda^- \approx 0$ and $e_\tau^{++}
\lambda^+\Pi_\sigma \lambda^- - e_\sigma^{++} \lambda^+ L
\lambda^- \approx 0$. Due to the nondegeneracy of the zweibein,
Eq.~(\ref{det(e)}), the solution to these two equations is
trivial, {\it i.e.} it implies $\lambda^+ L \lambda^- \approx 0$
and $\lambda^+\Pi_\sigma \lambda^-  \approx 0$, the last of which
is just the secondary constraint (\ref{phi0}).

To solve this system of equations for the Lagrange multipliers
and thus to describe explicitly the first class constraints, we
can use the auxiliary spinor fields $u_\alpha^I(\xi)$ defined as
in (\ref{cond1}), (\ref{cond2}). The general solution of
Eqs.~(\ref{dtP})--(\ref{dtP0}) obtained in such a framework can be
found in Appendix B (Eqs.~(\ref{Lab=A})--(\ref{l0=A})).
Schematically, it reads
\begin{eqnarray}
\label{Lab=}
L^{\alpha\beta} &= & b_{IJ}u^{\alpha I} u^{\beta J} + e^{++}_{\tau}
(\ldots) +  e^{--}_{\tau} (\ldots) \; ,
\\
\label{xi=} \xi^{\alpha} &=& \kappa_I \, u^{\alpha I} +
\frac{e^{++}_{\tau}}{e^{++}_{\sigma}}
(\partial_{\sigma} \theta \lambda^-) \lambda^{+ \alpha} -
\frac{e^{--}_{\tau}}{e^{--}_{\sigma}} (\partial_{\sigma} \theta
\lambda^+) \lambda^{- \alpha}
\; ,
\nonumber \\ {}
\\
 \label{l+=} l^+_{\alpha} &= & \omega^{(0)}\lambda_{\alpha}^{+} +
e^{++}_{\tau}
(\ldots) +  e^{--}_{\tau} (\ldots) \; ,
\\
\label{l-=}  l^-_{\alpha} &=& -\omega^{(0)}\lambda_{\alpha}^{-} +
e^{++}_{\tau}
(\ldots) +  e^{--}_{\tau} (\ldots)  \; ,
\\
\label{Lpmpm=} L^{\pm\pm} &=& \partial_{\sigma}e^{\pm \pm}_{\tau}
\pm 2e^{\pm \pm}_{\sigma} \omega^{(0)}
+ e^{++}_\tau  (\ldots )+ e^{--}_\tau  (\ldots ) \; ,
\\
\label{L(n)=} L^{(n)} &=& - 4 \mathrm{det}(e_m^a) \equiv -2
(e_\tau^{--}e_\sigma^{++} - e_\tau^{++}e_\sigma^{--} ) \; ,
\\
\label{l0=}  l^{(0)}&=& 0\; .
\end{eqnarray}
In this solution the parameters
\begin{eqnarray}
\label{paramb}
& bosonic\, : &  b^{IJ}=b^{JI}\, , \; \omega^{(0)} \, , \;
e_\tau^{\pm\pm}\, , \quad h^{\pm\pm}\, , \quad
\\ \label{paramf}
& fermionic \, : & \qquad  \kappa_{I}\; , \qquad
\end{eqnarray}
are indefinite. They correspond to the first class constraints
\begin{eqnarray}\label{PIJ}
{\cal P}^{IJ}& :=& {\cal P}_{\alpha\beta} u^{\alpha I} u^{\beta J}
\approx  0 \; ,
\\ \label{DI}
{\cal D}^I &:=& {\cal D}_\alpha u^{\alpha I} \approx 0 \; ,
\\ \label{G0}
G^{(0)} &:=&  \lambda^+_\alpha P^{\alpha (\lambda)}_+
- \lambda^-_\alpha P^{\alpha (\lambda)}_-
+ 2 e_\sigma^{++} P^\sigma_{++} \nonumber \\
&&- 2 e_\sigma^{--} P^\sigma_{--} \approx 0 \; ,
\end{eqnarray}
\begin{eqnarray}
\label{phi11} \tilde{\Phi}_{++} &:=& {\Phi}_{++}
+\partial_{\sigma} P_{++}^{\sigma}
 -2\Omega_{\sigma}^{(0)} P_{++}^{\sigma} -2e^{--}_{\sigma} {\cal N}
 \nonumber \\ &&
-\big[
\lambda^{-\alpha}\lambda^{-\beta} + \frac{2}{e^{++}_{\sigma}}
\big(\lambda^-_{\gamma} \Pi_{\sigma}^{\gamma \alpha}
\lambda^{+\beta}
 \nonumber \\ &&
-(\lambda^-\Pi_{\sigma}\lambda^+)\lambda^{-\alpha}\lambda^{+\beta}
+(\lambda^-\Pi_{\sigma}\lambda^-)\lambda^{+\alpha}\lambda^{+\beta}
\big) \big] {\cal P}_{\alpha \beta}  \nonumber \\ &&
-\frac{1}{e^{++}_{\sigma}}(\partial_{\sigma}\theta
\lambda^-)(\lambda^{+\alpha} {\cal D_{\alpha}}) \nonumber \\ &&
-\frac{1}{2e^{++}_{\sigma}} \big[e^{--}_{\sigma}
\Omega^{++}_{\sigma} + e^{++}_{\sigma} \Omega^{--}_{\sigma} +
i\partial_{\sigma}\theta \lambda^+\partial_{\sigma}\theta
\lambda^- \nonumber \\ && +\Pi^{\alpha\beta}_\sigma
(\partial_{\sigma}\lambda^+_{\alpha} \lambda^-_{\beta} -
\lambda^+_{\alpha}
\partial_{\sigma}\lambda^-_{\beta}) \big] \times \nonumber \\ &&
\times \left( \frac{\lambda_{\alpha}^- P_+^{\alpha
(\lambda)}}{e^{--}_{\sigma}} + \frac{\lambda_{\alpha}^+
P_-^{\alpha (\lambda)}}{e^{++}_{\sigma}} \right)
  \nonumber \\ &&
-\frac{1}{e^{++}_{\sigma}}
 (\partial_{\sigma} \lambda^-_{\alpha} +
 \Omega_{\sigma}^{(0)}\lambda_{\alpha}^-)P_-^{\alpha(\lambda)} \; ,
\end{eqnarray}
\begin{eqnarray}
\label{phi21}
\tilde{\Phi}_{--} &:=& {\Phi}_{--}
-\partial_{\sigma} P_{--}^{\sigma}
 +2\Omega_{\sigma}^{(0)} P_{--}^{\sigma} -2e^{++}_{\sigma} {\cal N}
 \nonumber \\ &&
+\big[
\lambda^{+\alpha}\lambda^{+\beta} - \frac{2}{e^{--}_{\sigma}}
\big(\lambda^+_{\gamma} \Pi_{\sigma}^{\gamma \alpha}
\lambda^{-\beta}
 \nonumber \\ &&
-(\lambda^+\Pi_{\sigma}\lambda^+)\lambda^{-\alpha}\lambda^{-\beta}
+(\lambda^+\Pi_{\sigma}\lambda^-)\lambda^{+\alpha}\lambda^{-\beta}
\big) \big] {\cal P}_{\alpha \beta}  \nonumber \\ &&
-\frac{1}{e^{--}_{\sigma}}(\partial_{\sigma}\theta
\lambda^+)(\lambda^{-\alpha} {\cal D_{\alpha}}) \nonumber \\ &&
+\frac{1}{2e^{--}_{\sigma}} \big[-e^{--}_{\sigma}
\Omega^{++}_{\sigma} - e^{++}_{\sigma} \Omega^{--}_{\sigma} +
i\partial_{\sigma}\theta \lambda^+\partial_{\sigma}\theta
\lambda^- \nonumber \\ && -\Pi^{\alpha\beta}_\sigma
(\partial_{\sigma}\lambda^+_{\alpha} \lambda^-_{\beta} -
\lambda^+_{\alpha}
\partial_{\sigma}\lambda^-_{\beta}) \big] \times \nonumber \\ &&
\times \left( \frac{\lambda_{\alpha}^- P_+^{\alpha
(\lambda)}}{e^{--}_{\sigma}} + \frac{\lambda_{\alpha}^+
P_-^{\alpha (\lambda)}}{e^{++}_{\sigma}} \right)
 \nonumber \\ &&
+\frac{1}{e^{--}_{\sigma}}
 (\partial_{\sigma} \lambda^+_{\alpha} +
\Omega_{\sigma}^{(0)}\lambda_{\alpha}^+)P_+^{\alpha(\lambda)}
 \; ,
\end{eqnarray}
and
\begin{eqnarray}\label{Ptau1}
P_{\pm \pm}^{\tau}
&\approx& 0 \; .
\end{eqnarray}
In Eqs.~(\ref{phi11}), (\ref{phi21}) ({\it cf.} Eqs. (\ref{vlpm}))
\begin{eqnarray}\label{Ompm=}
\Omega_\sigma^{++} &:=& \partial_{\sigma} \lambda^+C \lambda^+\; ,
\qquad \Omega_\sigma^{--} :=\partial_{\sigma} \lambda^-C \lambda^-
\\ \label{Om0=} \Omega_\sigma^{(0)} &:=&
\frac{1}{2}(\partial_{\sigma} \lambda^+C \lambda^- -
\lambda^+C \partial_{\sigma} \lambda^-) \; .
\end{eqnarray}
and the relation
\begin{eqnarray}\label{I=}
\delta_\alpha{}^\beta \approx \lambda^+_\alpha\lambda^{-\beta} -
\lambda^-_\alpha\lambda^{+\beta} - u^I_\alpha u^{J\beta}C_{IJ} \; ,
\\ \label{CL=}
\lambda^{\pm\beta}:= C^{\beta\alpha}\lambda^\pm_\alpha\; ,
\quad u^{I\beta}:= C^{\beta\alpha}u^I_\alpha\; ,
\end{eqnarray}
is used to remove the auxiliary variables $u^I_\alpha$ in all
places where it is possible. Note that (\ref{I=}) is a consequence
of the constraint (\ref{cl+l-}) and of the definition of the
$u^I_\alpha$ spinors, Eqs.~(\ref{cond1}), (\ref{cond2}) (see
further discussion on the use of $u$ variables below). Thus we are
allowed to use them in the solution of the equation for the Lagrange
multipliers and, then, in the definition of the first class
constraints, as the product of any two constraints is a first
class one since its Poisson brackets with any other constraint
vanishes weakly.

Using the Poisson brackets (\ref{canonical}), the first class
constraints generate gauge symmetries. In our dynamical system the
fermionic first class constraints (\ref{DI}) are the generators of
the $(n-2)$--parametric $\kappa$--symmetry
(\ref{kappa1})--(\ref{kappa3}). The ${\cal P}^{IJ}$ in
Eq.~(\ref{PIJ}) are the ${1 \over 2}(n-1)(n-2)$ generators of the
$b$-symmetry (\ref{b4}). The constraint $G^{(0)}$ (\ref{G0})
generates the $SO(1,1)$ gauge symmetry (\ref{SO(1,1)}).  Finally,
the constraints $\tilde{\Phi}_{\pm\pm}$, Eqs.~(\ref{phi11}),
(\ref{phi21}), generate worldvolume reparametrizations. They
provide a counterpart of the Virasoro constraints characteristic
of the Green--Schwarz superstring action. Thus, as it could be
expected, our $\Sigma^{(\frac{n(n+1)}2|n)}$ supersymmetric string is a
two-dimensional conformal field theory.

As it was noted above, the
presence of the first class constraints
 (\ref{Ptau1}) indicates the pure gauge nature of the fields
$e_\tau^{\pm\pm}(\xi)$; the freedom of the gauge fixing is,
nevertheless, restricted by the `topological' conditions
(\ref{det(e)}).

Note that the $\kappa$--symmetry and $b$--symmetry generators,
Eqs.~(\ref{DI}) and (\ref{PIJ}), are the $u_{\alpha}^I$ and
 $u_{\alpha}^I u_{\beta}^J$ components of
Eq.~(\ref{D})  and Eq.~(\ref{PX}), respectively, while all other
first class constraints can be defined without any reference to
auxiliary variables.

The use of the auxiliary spinors $u_\alpha^I(\xi)$ to define the
first class constraints requires some discussion. Clearly, any
spinor can be decomposed in the basis (\ref{basis}), but the use
of $u_\alpha^I$ to define constraints requires, to be rigorous, to
consider them as (auxiliary) dynamical variables, to introduce
momenta, and to take into account any additional constraints for
them, including Eqs.~(\ref{cond2}) and the vanishing of the
momenta conjugate to $u_\alpha^I$ ({\it cf.}~\cite{BZstr}). An
alternative is to consider these auxiliary spinors as defined by
(\ref{cond1}), (\ref{cond2}) and by the gauge symmetries of these
constraints, {\it i.e.} to treat them as some implicit functions
of $\lambda^\pm_\alpha$ ({\it cf.}~\cite{Grassi}). Such a
description can be obtained rigorously by the successive gauge
fixings of all the additional gauge symmetries that act only on
$u_\alpha^I$ and by introducing Dirac brackets accounting for all
the second class constraints for the $u_\alpha^I$ variables.
Nevertheless, with some precautions, the above simpler alternative
can be used from the beginning. In this case, one has to keep in
mind, in particular, that the $u_\alpha^I$'s do not commute with
$P_{\pm}^{\alpha (\lambda)}$. Indeed, as conditions (\ref{cond1})
have to be treated in a strong sense, one has to assume
$[P_{\pm}^{\alpha (\lambda)}(\sigma), u^I_\beta (\sigma^\prime
)]_P \approx \pm \lambda^\pm_\beta C^{\alpha\gamma} u_\gamma^I
\delta (\sigma - \sigma^\prime)$. However, one notices that this
does not change the result of the analysis of the number of first
and second class constraints among Eqs.~(\ref{PX})--(\ref{cl+l-}),
(\ref{phipm1}), (\ref{phi0}), which do not involve
$u_\gamma^I(\xi)$. The reason is that one only uses
$u_\gamma^I(\xi)$ as multipliers needed to extract the first and
second class constraints from the mixed ones (\ref{PX}),
(\ref{D}). Thus, the Poisson brackets of the projected constraints
${\cal P}_{\alpha\beta} u^{\alpha I} u^{\beta J}$, ${\cal
D}_\alpha u^{\alpha I}$ with other constraints ({\it e.g.},
$[{\cal P}_{\alpha\beta} u^{\alpha I} u^{\beta J}\, , \ldots ]_P$)
and the projected Poisson brackets of the original constraints
${\cal P}_{\alpha\beta}$, ${\cal D}_\alpha$ with the same ones
({\it e.g.}, $u^{\alpha I} u^{\beta J} \,[{\cal P}_{\alpha\beta}
\, , \ldots ]_P$) are equivalent in the sense that a non-zero
difference ($[{\cal P}_{\alpha\beta} u^{\alpha I} u^{\beta J}\, ,
\ldots ]_P\;-\, u^{\alpha I} u^{\beta J} \,[{\cal P}_{\alpha\beta}
\, , \ldots ]_P$) will be proportional to ${\cal P}_{\alpha\beta}$
or ${\cal D}_\alpha$ and, hence, will vanish weakly. This
observation allows us to use the basis (\ref{basis}) to solve the
equations (\ref{dtP})--(\ref{dtCll}), that is to say, to decompose
the constraints (\ref{PX})--(\ref{cl+l-}), (\ref{phipm1}),
(\ref{phi0}) into first and second class ones, without introducing
momenta for the $u_\gamma^I(\xi)$ and without studying the
constraints restricting these variables.

\bigskip

The remaining constraints are second class. In particular, these
are the $\lambda^{\pm}$ components of the fermionic constraints
(\ref{D}),
\begin{equation} \label{Dpm}
{\cal D}^\pm = {\cal D}_\alpha \lambda^{\alpha\pm} = \pi_\alpha
\lambda^{\alpha\pm} + i e_\sigma^{\pm\pm} \theta^\beta
\lambda_\beta^\mp \approx 0 \;
\end{equation}
with Poisson brackets
\begin{eqnarray}\label{DpmDpm}
\{ {\cal D}^+(\sigma) , {\cal D}^+ (\sigma^\prime )\}_P
&\approx& +2i e_\sigma^{++}  \delta (\sigma -\sigma^\prime )\; ,
\nonumber \\
\{ {\cal D}^+(\sigma) , {\cal D}^+ (\sigma^\prime )\}_P
&\approx& -2i e_\sigma^{--}  \delta (\sigma -\sigma^\prime )\; ,
\nonumber \\
\{ {\cal D}^+(\sigma) , {\cal D}^+ (\sigma^\prime )\}_P &\approx& 0 \;
\end{eqnarray}
(recall that, having in mind the possibility of fixing the
conformal gauge (\ref{cg}), we assume nondegeneracy of
$e_\sigma^{\pm\pm}(\sigma)$, {\it i.e.} that the expression
$1/e_\sigma^{\pm\pm}(\sigma)$ is well defined). The selection of
the basic second class constraints and the simplification of their
Poisson bracket algebra is a technically involved problem.

In the next section we show that the dynamical degrees of freedom
of our supersymmetric string in $\Sigma^{({n(n+1)\over 2}|n)}$, 
may be
presented in a more economic way in terms of constrained
$OSp(2n|1)$ supertwistors. The Hamiltonian mechanics also
simplifies in this symplectic supertwistor formulation. In
particular, all the first class constraints can be extracted
without using the auxiliary fields $u_\alpha^I$. The reason is
that the supertwistor variables are invariant under both
$\kappa$-- and $b$--symmetry. Thus, moving to the twistor form of
our action means rewriting it in terms of trivially $\kappa$-- and
$b$--invariant quantities, effectively removing all variables that
transform nontrivially under these gauge symmetries. Since the
description of $\kappa$-- and $b$--symmetries is the one requiring
the introduction of the $u_\alpha^I(\xi)$ fields, it is natural
that these are not needed in the supertwistor Hamiltonian
approach.

This consideration already allows us to calculate the number of
the (field theoretical worldsheet) degrees of freedom of our
$\Sigma^{(n(n+1)|n)}$ supersymmetric string model. 
The dynamical system described by the action
(\ref{St}) possesses ${1 \over 2}(n-1)(n-2)+5$ bosonic first class
constraints (equations (\ref{PIJ}), (\ref{G0}), (\ref{phi11}),
(\ref{phi21}) and (\ref{Ptau1})) out of a total number of $\frac12
n(n+1)+2n+8$ constraints (Eqs.~(\ref{PX}), (\ref{Plam}),
(\ref{Psig}), (\ref{Ptau}), (\ref{cl+l-}), (\ref{phipm1}) and
(\ref{phi0})). This leaves $4n+2$ bosonic second class
constraints. Since the phase space dimension corresponding to the
worldvolume bosonic fields ${\cal Z}^{\cal M}(\tau,
\sigma)=(X^{\alpha \beta}, \lambda^{\pm}_{\alpha},
e^{\pm\pm}_{\sigma}, e^{\pm \pm}_{\tau})$ is $2({n(n+1)\over 2}
+2n+4)$, the action (\ref{St}) turns out to have $(4n-6)$ bosonic
degrees of freedom.

Likewise, the $(n-2)$ fermionic first class constraints (\ref{DI})
and the $2$ fermionic second class constraints, Eqs.~(\ref{Dpm}),
reduce the original $2n$ phase space fermionic degrees of freedom
of the action (\ref{St}) down to $2$.

Thus our supersymmetric string model in $\Sigma^{({n(n+1)\over
2}|n)}$ superspace carries $(4n-6)$ bosonic and $2$ fermionic
worldvolume field theoretical degrees of freedom. Treating  the
number $n$ as the number of components of an irreducible spinor
representation of the $D$--dimensional Lorentz group $SO(1,D-1)$,
one finds

\bigskip
\begin{center}
\begin{tabular}
{|c|c|c|c|c|} \hline  $D$  &  $n$ & \, $\#_{bosonic \; d.o.f.} $ &
\, $\#_{fermionic \; d.o.f.}$ & \, BPS \; \cr
 &   & $ = 4n-6 $ &  $ =2$ & \,
states \; \cr \hline  3 & 2 & 2 & 2 & NO \cr \hline 4 & 4 & 10 & 2
& $1/2$ \cr \hline  6 & 8 & 26 & 2 & $6/8$ \cr \hline  10 & 16 &
58 & 2 & $14/16$ \cr \hline  11 & 32 & 122 & 2 & $30/32$ \cr
\hline
\end{tabular}
\end{center}

\bigskip

\noindent Thus, the number of bosonic degrees of freedom of our
$\Sigma^{({n(n+1)\over 2}|n)}$ supersymmetric string model exceeds that of
the Green--Schwarz superstring (where it exists, $4n-6 > 2(D-2)$),
while  the number of fermionic dimensions, $2$, is smaller than
that of the Green--Schwarz superstring for $D=6,10$.

The additional bosonic degrees of freedom might be treated as
higher spin degrees of freedom and/or as corresponding to the
additional `brane' central charges in the maximal supersymmetry
algebra (\ref{QQP}). The smaller number of physical fermionic
degrees of freedom just reflects the presence of supernumerary
$\kappa$--symmetries ($(n-2) > n/2$ for $n>4$) in our
$\Sigma^{(36|8)}$, and $\Sigma^{(528|32)}$, supersymmetric string models.
Our $\Sigma^{({n(n+1)\over 2}|n)}$ superstring model describes, as
argued, the excitations of a BPS state preserving $k=(n-2)$
supersymmetries (a ${30\over 32}$ BPS state for the 
supersymmetric string in the enlarged $D=11$ superspace 
$\Sigma^{(528|32)}$).

The search for solitonic solutions of the usual $D= 11$ and $D=10$
type II  supergravities with such properties is being carried out
at present \cite{Duff03,Hull03}. If successful, it would be
interesting to study how the additional bosonic degrees of freedom
of our model are mapped into the moduli of these solutions,
presumably related to the gauge fields of the supergravity
multiplet ({\it cf.}~\cite{JdA00}). Nevertheless, if it were shown
that such solutions do not appear in the standard $D=11$
supergravity, this could indicate that M-theory does require an
extension of the usual superspace for its adequate description.

To conclude this section we comment on the BPS preon
interpretation of our model. In accordance with \cite{BPS01}, it
can be treated as a composite of $\#_p=n-k=2$ BPS preons. To
support this conclusion one can have a look at the constraint
(\ref{PX}). As we have shown, it is a mixture of first and second
class constraints. However, performing a `conversion' of the
second class constraints \cite{conversion} to obtain first class
constraints (in a way similar to the one carried out for a
point--like model in \cite{BLS99}), one arrives at the first class
constraint
\begin{eqnarray} \label{I-PX}
{\cal P}_{\alpha\beta} & = &  P_{\alpha\beta} +
e_{\sigma}^{++} \tilde{\lambda}^-_\alpha \tilde{\lambda}^-_\beta -
e_{\sigma}^{--} \tilde{\lambda}^+_\alpha \tilde{\lambda}^+_\beta
\approx 0 \; ,
\end{eqnarray}
where the $\tilde{\lambda}^\pm_\alpha$ are related, but not just
equal, to $\lambda^\pm_\alpha$. In the quantum theory this first
class constraint can be imposed on quantum states giving rise to a
relation similar to Eq.~(\ref{kBPSdef}) with $\#_p=2$.

\section{Orthosymplectic twistor form of the 
$\Sigma^{({n(n+1)\over 2}|n)}$ 
 supersymmetric string action} \label{twistor}

A further analysis of the Hamiltonian mechanics of our supersymmetric string
model would become quite involved. Instead, we present in this
section a more economic description of our $\Sigma^{({n(n+1)\over
2}|n)}$ supersymmetric string model.

The action (\ref{St}) can be rewritten $(\alpha^\prime=1)$ in the
form
\begin{eqnarray}\label{St1}
S &=&  \int_{W^2}  [e^{++} \wedge (d\mu^{-\alpha}
\lambda^-_{\alpha} - \mu^{-\alpha} d\lambda^-_{\alpha} - i d\eta^-
\eta^-) \hspace{-1cm} \nonumber
\\
& -& e^{--} \wedge (d\mu^{+\alpha}  \lambda^+_{\alpha} -
\mu^{+\alpha}  d\lambda^+_{\alpha} - i d\eta^+ \eta^+) \nonumber
\hspace{-1cm}
\\
& - & \, e^{++} \wedge e^{--}] \; ,\hspace{-1cm}
\end{eqnarray}
where the bosonic $\mu^{\pm\alpha}$ and the fermionic $\eta^{\pm}$
are defined by
\begin{equation}
\label{mu+}  \mu^{\pm\alpha} = X^{\alpha\beta}
\lambda^{\pm}_{\beta}- {i\over 2} \theta^{\alpha} \theta^{\beta}
\lambda^{\pm}_{\beta} \; , \quad \eta^{\pm} = \theta^{\beta}
\lambda^{\pm}_{\beta} \; .
\end{equation}
Eqs.~(\ref{mu+}) are reminiscent of the Ferber generalization
\cite{Ferber} of the Penrose correspondence relation \cite{Pen}
(see also \cite{BL98,BPS01}). The two sets of $2n+1$ variables
belonging to the same real one-dimensional (Majorana--Weyl spinor)
representation of the worldsheet Lorentz group $SO(1,1)$,
\begin{eqnarray} \label{Ypm}
(\mu^{+\alpha}, \lambda_{\alpha}^{+}, \eta^+) := Y^{+\Sigma} \;\;
, \;\;  (\mu^{-\alpha}, \lambda_{\alpha}^{-}, \eta^-) :=
Y^{-\Sigma}\; , \quad
\end{eqnarray}
 can be treated as the components
of two $OSp(2n|1)$ supertwistors, $Y^{+ \Sigma}$ and $Y^{-
\Sigma}$. However, Eqs.~(\ref{mu+}) considered together imply the
following constraint:
\begin{eqnarray}\label{Con}
\lambda^+_{\alpha}\mu^{-\alpha} - \lambda^-_{\alpha} \mu^{+\alpha}
 - i \eta^- \eta^+ =0 \; .
\end{eqnarray}
One has to consider as well the `kinematic' constraint
(\ref{l+l-}), which breaks $GL(n,\mathbb{R})$ down to $Sp(n)$. In
terms of the supertwistors $Y^{\pm \Sigma}$ the action (\ref{St})
and the constraints (\ref{Con}), (\ref{l+l-})
can be written as
follows
\begin{eqnarray}\label{St002}
S &=& \int_{W^2} [e^{++} \wedge
dY^{-\Sigma}\, \Omega_{\Sigma\Pi} Y^{-\Pi}\,   \nonumber \\
&& - e^{--} \wedge dY^{+\Sigma}\, \Omega_{\Sigma\Pi} Y^{+\Pi}\ -
\, e^{++} \wedge e^{--}] \; ; \qquad
\end{eqnarray}
\begin{eqnarray}
\label{Con00}   Y^{+\Sigma}\, \Omega_{\Sigma\Pi}
Y^{-\Pi} & = & 0 \; , \qquad
\\ \label{l+l-00}
 Y^{+\Sigma}\, C_{\Sigma\Pi} Y^{-\Pi} & = & 1 \; , \qquad
\end{eqnarray}
where the nondegenerate matrix $\Omega_{\Sigma\Pi} =
-(-1)^{\mathrm{deg}({\pm\Sigma}) \mathrm{deg}({\pm\Pi})}
\Omega_{\Pi \Sigma}$ is the orthosymplectic metric,
\begin{eqnarray}
\label{OmLP} \Omega_{\Sigma\Pi} = \left\{ \begin{matrix} 0 &
\delta_\alpha{}^\beta & 0 \cr - \delta_\beta{}^\alpha & 0 & 0 \cr
0 & 0 & -i \end{matrix} \right\} \quad ,
\end{eqnarray}
preserved by $OSp(2n|1)$. The degenerate matrix $C_{\Sigma\Pi}$ in
Eq.~(\ref{l+l-00}) has the form
\begin{eqnarray} \label{CLP}
 C_{\Sigma\Pi} = \left\{ \begin{matrix} 0 & 0 & 0 \cr
0 & C^{\alpha\beta} & 0 \cr 0 & 0 & 0 \end{matrix} \right\} \quad
\end{eqnarray}
with $C^{\alpha\beta}$ defined in (\ref{Cab}).

One can also find the orthosymplectic twistor form for the action
(\ref{St1}) with unconstrained spinors. It reads
\begin{eqnarray}\label{St3}
S &=& \int_{W^2}  [e^{++} \wedge (d{\cal M}^{-\alpha}
\Lambda^-_{\alpha} - {\cal M}^{-\alpha}  d\Lambda^-_{\alpha} - i
d\chi^- \chi^-)   \hspace{-1cm} \nonumber
\\
& - &e^{--} \wedge (d{\cal M}^{+\alpha}  \Lambda^+_{\alpha} -
{\cal M}^{+\alpha}  d\Lambda^+_{\alpha} - i d\chi^+ \chi^+)
\nonumber \hspace{-1cm}
\\
& - & e^{++} \wedge e^{--} (C^{\alpha\beta}
\Lambda^+_{\alpha}\Lambda^-_{\beta})^2] \; ,
\end{eqnarray}
where
\begin{equation}\label{Mu+}
{\cal M}^{\pm\alpha} = X^{\alpha\beta} \Lambda^{\pm}_{\beta}-
{i\over 2} \theta^{\alpha} \theta^{\beta} \Lambda^{\pm}_{\beta} \;
, \quad \chi^{\pm} = \theta^{\beta} \Lambda^{\pm}_{\beta} \; .
\end{equation}
Eq.~(\ref{Mu+}) differs from (\ref{mu+}) only by replacement of
the constrained dimensionless $\lambda^\pm$ by the unconstrained
dimensionful  $\Lambda^\pm$. But, as a result, the $OSp(2n|1)$
supertwistors
\begin{eqnarray}
\label{Ups} && \Upsilon^{\pm \Sigma} := ({\cal M}^{\pm\alpha},
\Lambda^{\pm}_{\alpha}, \chi^{\pm})\; ,
\end{eqnarray}
 are restricted by only one condition similar to (\ref{Con00}),
\begin{eqnarray}
\label{Con001} && \Upsilon^{+\Sigma}\, \Omega_{\Sigma\Pi}
\Upsilon^{-\Pi} =0 \; .
\end{eqnarray}
The action in terms of $\Upsilon^{\pm\Sigma}$ includes the
degenerate matrix $C_{\Sigma\Pi}$, and reads
\begin{eqnarray}\label{St003}
S &=& \int_{W^2} [e^{++} \wedge d\Upsilon^{-\Sigma}\,
\Omega_{\Sigma\Pi} \Upsilon^{-\Pi}\,   \nonumber \\ && - e^{--}
\wedge d\Upsilon^{+\Sigma}\, \Omega_{\Sigma\Pi} \Upsilon^{+\Pi}\
\nonumber \\ && -\, e^{++} \wedge e^{--} \, (\Upsilon^{+\Sigma}\,
C_{\Sigma\Pi} \Upsilon^{-\Pi})^2] \; .
\end{eqnarray}

The global symmetry of our $\Sigma^{({n(n+1)\over 2}|n)}$
supersymmetric string is transparent now. The orthosymplectic supertwistors
$\Upsilon^{\pm\Sigma}$ are both in the fundamental representation
of the $OSp(2n|1)$ supergroup. The constraints (\ref{Con00}) (or
(\ref{Con001})) are also $OSp(2n|1)$ invariant. However, condition
(\ref{l+l-00}) (or the last term in the action (\ref{St003}))
breaks the $OSp(2n|1)$ invariance down to the semidirect product
${\Sigma}^{({n(n+1)\over 2}|n)} \times\!\!\!\!\!\!\supset Sp(n)$
of $Sp(n)\subset Sp(2n)$ and the maximal superspace group
$\Sigma^{({n(n+1)\over 2}|n)}$ (see Appendix A).

Summarizing, our  $\Sigma^{({n(n+1)\over 2}|n)}$ supersymmetric string model
breaks the $OSp(2n|1)$ symmetry down to a generalization
${\Sigma}^{({n(n+1)\over 2}|n)} \times\!\!\!\!\!\!\!\supset Sp(n)$
of the Poincar\'e supergroup. In contrast, both the point--like
model in \cite{BL98} and the tensionless superbrane model of
\cite{B02} possess full $OSp(2n|1)$ symmetry. This is in agreement
with treating $OSp(2n|1)$ as a generalized superconformal group,
as the standard conformal and superconformal symmetry is broken in
any model with mass, tension or another dimensionful parameter.

\section{\bf Hamiltonian analysis in the $OSp(2n|1)$ supertwistor
formulation} \label{twisthamilton}

The Hamiltonian analysis simplifies in the supertwistor
formulation (\ref{St002}) of the action (\ref{St}) . This is due
to the fact that moving from (\ref{St}) to (\ref{St002}) reduces
essentially the number of fields involved in the model.

The Lagrangian of the action (\ref{St002}) reads
\begin{eqnarray}
{\cal L} & = &
(e^{++}_{\tau}
\partial_{\sigma} Y^{^{-\Sigma}}
 - e^{++}_{\sigma} \partial_{\tau}
Y^{^{-\Sigma}}) \Omega_{_{\Sigma \Pi}} Y^{^{-\Pi}} \nonumber
\\
 &- & (e^{--}_{\tau} \partial_{\sigma} Y^{^{+\Sigma}}
 - e^{--}_{\sigma} \partial_{\tau}
Y^{^{+\Sigma}}) \Omega_{_{\Sigma \Pi}} Y^{^{+\Pi}}
\nonumber
\\
& - & (e^{++}_{\tau} e^{--}_{\sigma} - e^{++}_{\sigma}
e^{--}_{\tau}) \; ,
\end{eqnarray}
and involves the $2(2n+1+2)=4n+6$ configuration space worldvolume
fields
\begin{equation}\label{cZtwist}
\widetilde{\cal Z}^{\tilde{\cal M}}  \equiv \widetilde{\cal
Z}^{\tilde{\cal M}}(\tau, \sigma) =  \left( Y^{^{\pm \Sigma}} \, ,
\, e^{\pm\pm}_\tau \, , \, e^{\pm\pm}_\sigma \right)\; .
\end{equation}
The calculation of their canonical  momenta
\begin{eqnarray}\label{cPtwist}
\widetilde{P}_{\tilde{\cal M}}
= (P_{_{\pm\Sigma}}\,,\, P_{\pm\pm}^{\tau} \,, \,
P_{\pm\pm}^{\sigma}) = {\partial {\cal L} \over
\partial
(\partial_\tau \widetilde{\cal Z}^{\tilde{\cal M}})}
\end{eqnarray}
provides the following set of primary constraints:
\begin{eqnarray}
\label{pY}
{\cal P}_{_{\pm\Sigma}} & = & P_{_{\pm
\Sigma}} \mp e_{\sigma}^{\mp \mp} \Omega_{_{\Sigma \Pi}}
Y^{^{\pm\Pi}} \approx  0 \; ,\\
\label{psigma3} P_{\pm \pm}^{\sigma} & \approx & 0 \;, \\
\label{tau3} P_{\pm
\pm}^{\tau} & \approx & 0 \; .
\end{eqnarray}
Conditions (\ref{Con00}), (\ref{l+l-00}) should also be taken into
account after all the Poisson brackets are calculated and, hence,
are also primary constraints,
\begin{eqnarray}
\label{U1}
{\cal U} &:= &
Y^{+\Sigma}\, \Omega_{\Sigma\Pi}
Y^{-\Pi} \approx 0 \; ,
\end{eqnarray}
\begin{eqnarray}\label{c2l+l-}
{\cal N} & := & Y^{+\Sigma}\, C_{\Sigma\Pi} Y^{-\Pi} -1 \approx 0
\; .
\end{eqnarray}

The {\it canonical} Hamiltonian density ${\cal H}_0$ corresponding
to the action (\ref{St002}), reads
\begin{eqnarray} \label{H2Y}
{\cal H}_0 = & [-e^{++}_{\tau}
\partial_{\sigma} Y^{^{-\Sigma}} \Omega_{_{\Sigma \Pi}} Y^{^{-\Pi}} +
e^{--}_{\tau} \partial_{\sigma} Y^{^{+\Sigma}} \Omega_{_{\Sigma
\Pi}}
Y^{^{+\Pi}} \nonumber \\
& + \, (e^{++}_{\tau} e^{--}_{\sigma} - e^{++}_{\sigma}
e^{--}_{\tau}) ] \; . \hspace{-1cm}
\end{eqnarray}
The  preservation of the primary constraints under
$\tau$--evolution (see Sec.~\ref{Hamiltonian}) leads to the
secondary constraints
\begin{eqnarray}
\label{Ph++}
\Phi_{++} &=& \partial_{\sigma} Y^{^{-\Sigma}}
\Omega_{_{\Sigma \Pi}} Y^{^{- \Pi}} - e^{--}_{\sigma}
\approx 0 \; ,
\\
\label{Ph--}
\Phi_{--} & =& \partial_{\sigma} Y^{^{+ \Sigma}}
\Omega_{_{\Sigma \Pi}} Y^{^{+ \Pi}} - e^{++}_{\sigma}
\approx 0 \; .
\\
\label{Ph0} \Phi^{(0)} &=& \partial_{\sigma} Y^{^{+ \Sigma}}
\Omega_{_{\Sigma \Pi}} Y^{^{- \Pi}} - Y^{^{+ \Sigma}}
\Omega_{_{\Sigma \Pi}}\partial_{\sigma}  Y^{^{- \Pi}} \approx 0 \;
. \qquad
\end{eqnarray}

Again (see Sec.~\ref{Hamiltonian}) the canonical Hamiltonian
vanishes on the surface of constraints (\ref{Ph++}), (\ref{Ph--}),
and thus the $\tau$--evolution is defined by the Hamiltonian
density ({\it cf.} (\ref{H=})) {\setlength\arraycolsep{2pt}
\begin{eqnarray}\label{H(Y)=}
{\cal H}^\prime &=& - e_\tau^{++} \Phi_{++} + e_\tau^{--}
\Phi_{--}  + l^{(0)}  \Phi^{(0)} + L^{\pm \Sigma} {\cal P}_{\pm
\Sigma} +
\nonumber \\
&+& L^{(0)} {\cal U} +  L^{(n)} {\cal N} + L^{\pm\pm}
P_{\pm\pm}^{\sigma} + h^{\pm\pm} P_{\pm\pm}^\tau  \;
\end{eqnarray}}
and the canonical Poisson brackets
\begin{eqnarray} \label{YPY}
&& [ P_{_{\pm\Lambda}}(\sigma)\, , \,  Y^{^{\pm\Sigma}}
(\sigma^{\prime})
\}_{_P} = - \delta^{^{\Sigma}}
_{_{\Lambda}} \;  \delta(\sigma-\sigma^{\prime}) , \quad  \\
\label{ePe1} && [ e_\sigma^{\pm\pm} (\sigma ) \, , \,
P_{{\pm\pm}}^{\sigma} (\sigma^{\prime}) ]_{_P} =
\delta(\sigma-\sigma^{\prime}) \; , \\
\label{ePe2} && [e_\tau^{\pm\pm} (\sigma ) \, , \,
P_{{\pm\pm}}^{\tau} (\sigma^{\prime}) ]_{_P} =
\delta(\sigma-\sigma^{\prime})
 \; , \quad
\end{eqnarray}

Then the $\tau$--preservation requirement of the primary and
secondary constraints results in the following system of equations
for the Lagrange multipliers
\begin{eqnarray}
\label{dtP+S}
L^{^{+\Sigma}}& \approx &
{e_\tau^{--} \over e_\sigma^{--} }\partial_{\sigma}  Y^{^{+\Sigma}}
+ {\partial_{\sigma} e_\tau^{--} - L^{--}
\over 2e_\sigma^{--} }Y^{^{+\Sigma}}  + \nonumber \\
&+& {l^{(0)} \over e_\sigma^{--} }\partial_{\sigma}  Y^{^{-\Sigma}}
 +
{\partial_{\sigma} l^{(0)} - L^{(0)}
\over 2e_\sigma^{--} }Y^{^{-\Sigma}} -
\nonumber \\
&-& {L^{(n)}
\over 2e_\sigma^{--} }Y^{^{-\Pi}} (C\Omega)_{_{\Pi}}{}^{^{\Sigma}} \; ,
\\ \label{dtP-S}
L^{^{-\Sigma}}& \approx &
{e_\tau^{++} \over e_\sigma^{++} }\partial_{\sigma}  Y^{^{-\Sigma}}
+ {\partial_{\sigma} e_\tau^{++} - L^{++}
\over 2e_\sigma^{++} }Y^{^{-\Sigma}}  - \nonumber \\
&-& {l^{(0)} \over e_\sigma^{++} }\partial_{\sigma}  Y^{^{+\Sigma}}
 -
{\partial_{\sigma} l^{(0)} + L^{(0)}
\over 2e_\sigma^{++} }Y^{^{+\Sigma}} -
\nonumber \\
&-& {L^{(n)}
\over 2e_\sigma^{++} }Y^{^{+\Pi}} (C\Omega)_{_{\Pi}}{}^{^{\Sigma}} \; ,
\\ \label{dtU(Y)}
L^{^{+\Sigma}}  \Omega_{_{\Sigma\Pi}} Y ^{^{-\Pi}}  &\approx&
L^{^{-\Sigma}} \Omega_{_{\Sigma\Pi}} Y ^{^{+\Pi}}   \; ,
\\
\label{dtN(Y)}
L^{^{+\Sigma}}  C_{_{\Sigma\Pi}} Y ^{^{-\Pi}}  & \approx &
L^{^{-\Sigma}} C_{_{\Sigma\Pi}} Y ^{^{+\Pi}}   \; ,
\\
\label{dtP++Y}
L^{^{-\Sigma}}  \Omega_{_{\Sigma\Pi}} Y ^{^{-\Pi}} & \approx &
e_\tau^{--} \; ,
\\
\label{dtP--Y}
L^{^{+\Sigma}}  \Omega_{_{\Sigma\Pi}} Y ^{^{+\Pi}} & \approx &
e_\tau^{++} \; ,
\end{eqnarray}
and
\begin{eqnarray}
\label{dtPh++Y}
&& L^{--}  \approx \partial_\sigma
L^{^{-\Sigma}}  \Omega_{_{\Sigma\Pi}} Y ^{^{-\Pi}}
- L^{^{-\Sigma}}  \Omega_{_{\Sigma\Pi}} \partial_\sigma Y ^{^{-\Pi}} \; ,
\\
\label{dtPh--Y}
&& L^{++}  \approx  \partial_\sigma
L^{^{+\Sigma}}  \Omega_{_{\Sigma\Pi}} Y ^{^{+\Pi}}
- L^{^{+\Sigma}}  \Omega_{_{\Sigma\Pi}} \partial_\sigma Y ^{^{+\Pi}} \; ,
\\ \label{dtPh0Y}
&& \sum\limits_{\pm}
\left(
\partial_\sigma
L^{^{\pm\Sigma}}  \Omega_{_{\Sigma\Pi}} Y ^{^{\mp\Sigma}}
- L^{^{\pm\Sigma}}  \Omega_{_{\Sigma\Pi}} \partial_\sigma Y ^{^{\mp\Sigma}}
\right) \approx 0 \; . \qquad
\end{eqnarray}
where $(C\Omega)_{_{\Pi}}{}^{^{\Sigma}}:= C_{_{\Pi\Lambda}}
\Omega^{^{\Lambda\Sigma}}$ and $\Omega^{^{\Sigma \Pi}} = -
\Omega_{_{\Sigma \Pi}}$ is the inverse of the orthosymplectic
metric (\ref{OmLP}),
\begin{equation}\label{Om-1}
\Omega_{_{\Sigma \Lambda}}  \Omega^{^{\Lambda \Pi}}=
\delta_{_{\Sigma}}^{^{\Pi}}\; , \quad \Omega^{^{\Sigma \Pi}} =
\left\{ \begin{matrix} 0 & - \delta_\beta{}^\alpha & 0 \cr
\delta_\alpha{}^\beta & 0 & 0 \cr 0 & 0 & i \end{matrix} \right\}
\quad .
\end{equation}

Equations (\ref{dtP+S})--(\ref{dtP--Y}) come from the preservation
of the primary constraints, while
Eqs.~(\ref{dtPh++Y})--(\ref{dtPh0Y}) from the preservation of the
secondary constraints. Again, as in Sec. IV, one can follow the
appearance of the secondary constraint (\ref{Ph0}) by considering
Eqs.~(\ref{dtP+S})--(\ref{dtP--Y}) with $l^{(0)}=0$.

Denoting
\begin{eqnarray}
\label{A0=}
A_\sigma^{(0)}& = &
{1\over 2} \left( \partial_\sigma Y^{^{+\Sigma}}
 C_{_{\Sigma\Pi}} Y^{^{-\Pi}}
- Y ^{^{+\Sigma}} C_{_{\Sigma\Pi}}
 \partial_\sigma Y^{^{-\Pi}} \right) \; ,
\qquad \\
\label{A++=}
A_\sigma^{++}& = & \partial_\sigma Y^{^{+\Sigma}}
C_{_{\Sigma\Pi}} Y^{^{+\Pi}}
\; ,
\\
\label{A--=}
A_\sigma^{--}& = &  \partial_\sigma Y^{^{-\Sigma}}
C_{_{\Sigma\Pi}} Y^{^{-\Pi}}
\; ,
\\
\label{B0=}
B^{(0)} &=& {\cal S}
-{\partial_\sigma Y^{+}\Omega\partial_\sigma Y^{-}
\over 2 e_\sigma^{++}e_\sigma^{--}} \; ,
\\ \label{calS=}
{\cal S} &=& {1\over 2} \left(
{A_\sigma^{++}\over e_\sigma^{++}} +
{A_\sigma^{--}\over e_\sigma^{--}}
\right) \; ,
\end{eqnarray}
one can write the general solution of
Eqs.~(\ref{dtP+S})--(\ref{dtP--Y}) in the form
\begin{eqnarray}
\label{L+S}
L^{^{+\Sigma}}& \approx & \omega^{(0)} Y^{^{+\Sigma}}
+ {e_\tau^{--}\over e_\sigma^{--} }
\left( \partial_{\sigma}  Y^{^{+\Sigma}}
- A_\sigma^{(0)}  Y^{^{+\Sigma}} - \right.   \nonumber \\
&& \left.  \qquad
- e_\sigma^{++} B^{(0)} Y^{^{-\Sigma}} +
  e_\sigma^{++} (Y^-C\Omega)^{^{\Sigma}}
\right) + \nonumber \\
&+& e_\tau^{++}
\left(  B^{(0)} Y^{^{-\Sigma}} -
(Y^-C\Omega)^{^{\Sigma}}\right)  \; ,
\\
\label{L-S}
L^{^{-\Sigma}}& \approx & - \omega^{(0)} Y^{^{-\Sigma}}
+ {e_\tau^{++}\over e_\sigma^{++} }
\left( \partial_{\sigma}  Y^{^{-\Sigma}}
+ A_\sigma^{(0)}  Y^{^{-\Sigma}} + \right.   \nonumber \\
&& \left.  \qquad
+ e_\sigma^{--} B^{(0)} Y^{^{+\Sigma}} -
 e_\sigma^{--} (Y^+C\Omega)^{^{\Sigma}}
\right) - \nonumber \\
&-& e_\tau^{--}
\left(  B^{(0)} Y^{^{+\Sigma}} -
(Y^+C\Omega)^{^{\Sigma}}\right)  \; ,
\\ \label{L0(Y)}
L^{(0)} &=&
2(e_\tau^{--}e_\sigma^{++} - e_\tau^{++}e_\sigma^{--}) \,
B^{(0)} \; ,
\\ \label{LpmpmY}
L^{\pm\pm}& = & \partial_{\sigma}e^{\pm \pm}_{\tau}\mp
2e^{\pm \pm}_\tau  A^{(0)}_\sigma
\pm 2e^{\pm \pm}_{\sigma} \omega^{(0)} \; ,
\\ \label{Ln(Y)}
L^{(n)} &=& - 4 \mathrm{det}(e_m^a) \equiv -2
(e_\tau^{--}e_\sigma^{++} - e_\tau^{++}e_\sigma^{--} ) \; , \quad
\\ \label{l0(Y)}
l^{(0)} &=& 0 \; .
\end{eqnarray}
Note that Eqs.~(\ref{Ln(Y)}), (\ref{l0(Y)}) have the same form as
Eqs.~(\ref{L(n)=A}), (\ref{l0=A}) (Appendix B) for the Lagrange
multipliers  in the original formulation, and Eqs.~(\ref{LpmpmY})
are similar to Eqs.~(\ref{Lpmpm=A}).

The above solution contains the indefinite worldsheet field
parameters $h^{\pm\pm}(\xi)$, $\omega^{(0)}(\xi)$ and
$e_\tau^{\pm\pm}(\xi)$ which correspond to the five first class
constraints which generate the gauge symmetries of the symplectic
twistor formulation of our $\Sigma^{({n(n+1)\over 2}|n)}$
supersymmetric string model. They are
\begin{eqnarray}\label{PtauY}
P_{\pm \pm}^{\tau}
&\approx& 0 \;
\end{eqnarray}
and
\begin{eqnarray}
\label{G0(Y)}
G^{(0)} &:=&
Y^{^{+\Sigma}}{\cal P}_{_{+\Sigma}} -
Y^{^{-\Sigma}} {\cal P}_{_{-\Sigma}} +
\nonumber \\ &+&
2e_\sigma^{++} P^\sigma_{++} - 2  e_\sigma^{--} P^\sigma_{--}
\approx 0 \, ,
\\
\label{IPh++Y}
\tilde{\Phi}_{++} &:=& {\Phi}_{++} +
\partial_{\sigma} P_{++}^{\sigma} + 2A_{\sigma}^{(0)}
P_{++}^{\sigma} + \nonumber \\ &+ & 2 e^{--}_{\sigma} B^{(0)}
{\cal U} - 2e^{--}_{\sigma} {\cal N} + {\cal
F}_{++}^{^{\pm\Sigma}} {\cal P}_{{\pm\Sigma}} \; ,
\\
\label{IPh--Y}
\tilde{\Phi}_{--} &:=& {\Phi}_{--} -
\partial_{\sigma} P_{--}^{\sigma} +
2A_{\sigma}^{(0)} P_{--}^{\sigma}
+ \nonumber \\ &+ & 2 e^{++}_{\sigma} B^{(0)} {\cal U}
- 2e^{++}_{\sigma} {\cal N} +
 {\cal F}_{--}^{^{\pm\Sigma}} {\cal P}_{{\pm\Sigma}}\; ,
\end{eqnarray}
where
\begin{eqnarray}
\label{cF+++S}
{\cal F}_{++}^{^{+\Sigma}} & = &
- B^{(0)} Y^{^{-\Sigma}}
+ (Y^-C\Omega)^{^{\Sigma}} \; ,
\\ \label{cF++-S}
{\cal F}_{++}^{^{-\Sigma}} & = & - {1\over e_\sigma^{++}}
[\partial_{\sigma} Y^{^{-\Sigma}} +  A_\sigma^{(0)} Y^{^{-\Sigma}}
+
\nonumber \\
&+& B^{(0)} e_\sigma^{--}
Y^{^{+\Sigma}} - e_\sigma^{--} (Y^+C\Omega)^{^{\Sigma}}] \; ,
\\
\label{cF--+S} {\cal F}_{--}^{^{+\Sigma}} & = & {1\over
e_\sigma^{--}} [\partial_{\sigma} Y^{^{+\Sigma}} - A_\sigma^{(0)}
Y^{^{+\Sigma}} -
\nonumber \\
&-& B^{(0)} e_\sigma^{++}
Y^{^{-\Sigma}} + e_\sigma^{++} (Y^-C\Omega)^{^{\Sigma}}]\; ,
\\
\label{cF---S}
{\cal F}_{--}^{^{-\Sigma}} & = &
-  B^{(0)} Y^{^{+\Sigma}}
+ (Y^+C\Omega)^{^{\Sigma}} \; .
\end{eqnarray}
Using Poisson brackets, the constraint (\ref{G0(Y)}) generates the
$SO(1,1)$ worldsheet Lorentz gauge symmetry, (\ref{IPh++Y}) and
(\ref{IPh--Y})  are the reparametrization (Virasoro) generators,
and the symmetry generated by Eqs.~(\ref{PtauY}) indicates the
pure gauge nature of the $e_\tau^{\pm\pm}(\xi)$ fields  (again,
subject to the nondegeneracy condition (\ref{det(e)}) that
restricts the gauge choice freedom for them).

Note that both the $b$--symmetry and the $\kappa$--symmetry
generators, Eqs.~(\ref{PIJ}) and (\ref{DI}), are not present in
the symplectic supertwistor formulation. Actually, the number of
variables in this formulation minus the constraint among them,
Eq.~(\ref{Con00}), is $(4n+6)-1$ and equal to the number of
variables in the previous formulation $(\frac{n(n+1)}2+n+2n+4)$,
minus the number of $b$-- and $\kappa$--symmetry generators
$(\frac{(n-1)(n-2)}2+(n-2))$. This clearly indicates that the
transition to the supertwistor form of the action corresponds to
an implicit gauge fixing of these symmetries and the removal of
the additional variables, since the remaining supertwistor ones
are invariant under both $b$-- and $\kappa$--symmetry
\footnote{These invariance was known for the massless
superparticle and the tensionless superstring cases, see {\it
e.g.} \cite{Sokatchev,BL98,BLS99,B02}; {\it cf.}~\cite{ZU02}.}.

Other constraints are second class. Indeed, {\it e.g.} the algebra
of the constraints ${\cal P}_{\pm\Sigma}$, Eq.~(\ref{pY}),
\begin{eqnarray} \label{cPcP+}
&& [ {\cal P}_{_{+\Sigma}} (\sigma ) \, , \, {\cal
P}_{_{+\Lambda}}(\sigma^{\prime}) \}_{_P} = 2 e_\sigma^{--}
\Omega_{_{\Lambda \Sigma}}
\delta(\sigma-\sigma^{\prime})\; ,  \\
\label{cPcP-} && [ {\cal P}_{_{-\Sigma}} (\sigma ) \, , \, {\cal
P}_{_{-\Lambda}}(\sigma^{\prime}) \}_{_P} = - 2 e_\sigma^{++}
\Omega_{_{\Lambda\Sigma}} \delta(\sigma-\sigma^{\prime})\; , \quad
\\
\label{cPcP+-} && [ {\cal P}_{_{+\Sigma}} (\sigma ) \, , \, {\cal
P}_{_{-\Lambda}}(\sigma^{\prime}) \}_{_P} = 0 \; ,
\end{eqnarray}
clearly indicates their second class nature. As so, one can
introduce the  graded Dirac (or starred \cite{Dirac}) brackets
that allows one to put them strongly equal to zero. For any
arbitrary two (bosonic or fermionic) functionals $f$ and $g$ of
the canonical variables (\ref{cZtwist}), (\ref{cPtwist}) they are
defined by
\begin{eqnarray} \label{PBY}
& [f(\sigma_1), g(\sigma_2)\}_{_D} = [f(\sigma_1) ,
g(\sigma_2)\}_{_P} - \nonumber \\
& -{1 \over 2} \int d\sigma \left(
\frac{1}{e^{--}_{\sigma}(\sigma)} [f(\sigma_1) , {\cal
P}_{_{+\Sigma}}(\sigma) \}_{_P}\Omega^{^{\Pi\Sigma}} [ {\cal
P}_{_{+\Pi}}(\sigma), g(\sigma_2)\}_{_P} \right. \nonumber
 \\
 & \left.  -{1 \over e^{++}_{\sigma}(\sigma)} [f(\sigma_1) ,
 {\cal P}_{_{-\Sigma}}(\sigma) \}_{_P}\Omega^{^{\Pi\Sigma }} [ {\cal
P}_{_{-\Pi}}(\sigma), g(\sigma_2)\}_{_P} \right) \, . \qquad
\end{eqnarray}

\noindent Using these and reducing further the number of phase
space degrees of freedom by setting $P_{_{\pm\Sigma}}=0$ strongly,
the supertwistor becomes a self-conjugate variable,
\begin{equation} \label{DB2}
 [Y^{^{\pm \Sigma}}(\sigma), Y^{^{\pm
\Pi}}(\sigma^{\prime})\}_{_D} = \mp {1 \over 2e^{\mp \mp}_{\sigma}} \,
\Omega^{^{\Sigma \Pi}} \, \delta (\sigma - \sigma^{\prime}) \; .
\end{equation}
For the `components' of the supertwistor, Eq.~(\ref{DB2}) implies
{\setlength\arraycolsep{2pt}
\begin{eqnarray} \label{DBbose}
[\lambda^{\pm}_{\alpha}(\sigma),
\mu^{\pm \beta}(\sigma^\prime)]_{_D} &=& \mp
{1 \over 2e^{\mp \mp}_{\sigma}}
\delta_{\alpha}^{\beta}  \delta(\sigma-\sigma^\prime) \ , \\
\label{DBfermi} \{\eta^{\pm}(\sigma),
\eta^{\pm}(\sigma^\prime)\}_{_D} &=& \mp
{i \over 2e^{\mp \mp}_{\sigma}} \
\delta(\sigma-\sigma^\prime) \; .
\end{eqnarray}}
The Dirac brackets for $e_\sigma^{\pm\pm}$, $e_\tau^{\pm\pm}$ and
$P^\tau_{\pm\pm}$ coincide with the Poisson brackets, while for
$P^\sigma_{\pm\pm}$ one finds
\begin{eqnarray}
\label{P++D} [P^\sigma_{++}(\sigma), ... ]_{_D} & =&
[P^\sigma_{++}(\sigma), ... ]_{_P} \nonumber \\ && - {1\over
2e_\sigma^{++}} Y^{^{-\Sigma}}(\sigma) [{\cal
P}_{_{-\Sigma}}(\sigma), ...  \}_{_P}\; ,  \qquad
\\
\label{P--D} [P^\sigma_{--}(\sigma), ... ]_{_D} &=&
[P^\sigma_{--}(\sigma), ... ]_{_P} - \nonumber \\ && - {1\over
2e_\sigma^{--}} Y^{^{+\Sigma}}(\sigma) [{\cal
P}_{_{+\Sigma}}(\sigma), ...  \}_{_P}\; . \qquad
\end{eqnarray}
However, $P^\sigma_{\pm\pm}(\sigma)$ still commute among
themselves, $[P^\sigma_{\pm\pm}(\sigma),
P^\sigma_{\pm\pm}(\sigma^\prime)]_{_D} =0=[P^\sigma_{++}(\sigma),
P^\sigma_{--}(\sigma^\prime)]_{_D}$.

When the constraints (\ref{pY}) are taken as strong equations, the
first class constraints (\ref{G0(Y)})--(\ref{IPh--Y}) simplify to
\begin{eqnarray}
\label{G0(Y)D}
G^{(0)} &:=&
2e_\sigma^{++} P^\sigma_{++} - 2  e_\sigma^{--} P^\sigma_{--}
\approx 0 \, ,
\\
\label{IPh++YD} \tilde{\Phi}_{++} &:=& {\Phi}_{++} +
\partial_{\sigma} P_{++}^{\sigma} + 2A_{\sigma}^{(0)} P_{++}^{\sigma}
+
\nonumber \\ &+&
2e^{--}_{\sigma} B^{(0)} {\cal U} - 2e^{--}_{\sigma} {\cal N} \; ,
\\
\label{IPh--YD} \tilde{\Phi}_{--} &:=& {\Phi}_{--} -
\partial_{\sigma} P_{--}^{\sigma} +
2A_{\sigma}^{(0)} P_{--}^{\sigma}
+ \nonumber \\ &+ & 2 e^{++}_{\sigma} B^{(0)} {\cal U}
- 2e^{++}_{\sigma} {\cal N} \; ,
\end{eqnarray}
and the remaining second class constraints can be taken in the form
\begin{eqnarray}
\label{K0}
K^{(0)} &:=&
e_\sigma^{++} P^\sigma_{++} +   e_\sigma^{--} P^\sigma_{--}
\approx 0 \, , \\
\label{N2} {\cal N} & = & Y^{+\Sigma}\,
C_{\Sigma\Pi} Y^{-\Pi} -1 \approx 0
\; , \\
\label{U2} {\cal U} &= & Y^{+\Sigma}\, \Omega_{\Sigma\Pi}Y^{-\Pi}
\approx 0 \; ,
\\
\label{Ph0II} \Phi^{(0)} &=& \partial_{\sigma} Y^{^{+ \Sigma}}
\Omega_{_{\Sigma \Pi}} Y^{^{- \Pi}} - Y^{^{+ \Sigma}}
\Omega_{_{\Sigma \Pi}}\partial_{\sigma}  Y^{^{- \Pi}} \approx 0 \;
. \qquad
\end{eqnarray}
One has to take into account that, under the Dirac brackets,
$P^\sigma_{\pm\pm}$ and $Y^{\mp\Sigma}$ do not commute,
\begin{eqnarray}
\label{P++Y} [P^\sigma_{++}(\sigma), Y^{-\Sigma} (\sigma^\prime)
]_{_D} & =& {1\over 2e_\sigma^{++}} Y^{^{-\Sigma}}(\sigma)
\delta(\sigma -  \sigma^\prime)\; ,  \qquad
\\
\label{P--Y} [P^\sigma_{--}(\sigma), Y^{+\Sigma} (\sigma^\prime)
]_{_D} & =& {1\over 2e_\sigma^{--}} Y^{^{+\Sigma}}(\sigma)
\delta(\sigma -  \sigma^\prime)\; .  \qquad
\end{eqnarray}
Then one easily checks that, under Dirac brackets, $G^{(0)}$
generates the $SO(1,1)$ transformations of the supertwistors,
\begin{eqnarray}
\label{[G0Y]} [G^{(0)}(\sigma), Y^{\pm \Sigma} (\sigma^\prime)
]_{_D} & =& \mp  Y^{^{\pm\Sigma}}(\sigma) \delta(\sigma -
\sigma^\prime)\; .  \qquad
\end{eqnarray}
On the other hand, one finds that the second class constraint
${\cal U}$ interchanges the $Y^{+\Sigma}$ and $Y^{-\Sigma}$
supertwistors,
\begin{eqnarray}
[\, {\cal U}(\sigma)\, , Y^{+\Sigma} (\sigma^\prime) ]_{_D} &=&
 \; {1\over 2e_\sigma^{--}} Y^{^{-\Sigma}}(\sigma)
\delta(\sigma -  \sigma^\prime)\; ,  \qquad  \nonumber
\\ \label{U0Y}
 [\, {\cal U}(\sigma), Y^{-\Sigma} (\sigma^\prime) ]_{_D} &=&
  {1\over 2e_\sigma^{++}} Y^{^{+\Sigma}}(\sigma)
\delta(\sigma -  \sigma^\prime)\; .  \qquad
\end{eqnarray}
It is interesting to note that in the original supertwistor
formulation of the $D=4$, $N=1$ superparticle \cite{Ferber} there
exists a counterpart of the ${\cal U}$ constraint; however, there
it is the first class constraint generating the internal $U(1)$
symmetry \footnote{See \cite{SG89} for a detailed study of the
Hamiltonian mechanics in the twistor--like formulation of the
$D=4$ superparticle, where the possibility of constraint class
transmutation was noted.}.

The Dirac brackets of the above second class constraints
(\ref{K0})--(\ref{Ph0II}) can be found in Appendix B,
Eqs.~(\ref{DB1})-(\ref{DB5}). They are characterized by the matrix

\vspace{5pt}
\begin{center}
\begin{tabular}
{|c|c|c|c|c|} \hline \tt $[... \downarrow \, , ... \rightarrow
\}_{_D} \approx$ & $(\Phi^{(0)}(\sigma^\prime)+$ & ${\cal U}
(\sigma^\prime)$ &  $ K^{(0)} (\sigma^\prime)$ & ${\cal N}
(\sigma^\prime) $ \cr
 & $+{\cal S} K^{(0)} (\sigma^\prime))$   &  &    & \cr
\hline \tt $(\Phi^{(0)}+ {\cal S}K^{(0)}) (\sigma)$  & 0 & $-
\delta_{\sigma\,\sigma^\prime}$ & 0 &  0 \cr \hline \tt ${\cal U}
(\sigma)$  & $\delta_{\sigma\,\sigma^\prime}$ & 0 & 0 & 0 \cr
\hline \tt $K^{(0)} (\sigma)$ & 0 & 0 & 0 &
$\delta_{\sigma\,\sigma^\prime}$\cr \hline \tt ${\cal N} (\sigma)$
& 0 & 0 & $- \delta_{\sigma\,\sigma^\prime}$ & 0  \cr \hline
\end{tabular}
\end{center}
\vspace{5pt}
 \noindent where ${\cal S}(\sigma)= {1\over 2} \left(
\frac{A_{\sigma}^{++}(\sigma)}{e^{++}_\sigma(\sigma)} +
\frac{A_{\sigma}^{--}(\sigma)}{e^{--}_\sigma(\sigma)} \right) $
(Eq.~(\ref{calS=})) and $\delta_{\sigma\, \sigma^\prime}\equiv
\delta(\sigma-\sigma^\prime)$. This table indicates that the
$K^{(0)}$ constraint is canonically conjugate to ${\cal N}$ while
the second class constraint $\Phi^{(0)} + {\cal S} K^{(0)}$ is
conjugate to ${\cal U}$. One may easily pass to the (doubly
starred) Dirac brackets with respect to the above mentioned four
second class constraints. However, the new Dirac brackets for the
supertwistor variables would have a very complicated form, so that
it looks more practical either to apply the formalism using
(singly starred) Dirac brackets (Eq.~(\ref{PBY})) and simple first
and second class constraints, Eqs.~(\ref{G0(Y)D})--(\ref{IPh--YD})
and (\ref{K0})--(\ref{Ph0II}), or to search for a conversion
\cite{conversion} of the remaining second  class constraints into
first class ones. Note that a phenomenon similar to conversion
occurs when one moves from (\ref{St002})  to the dynamical system
with unnormalized twistors described by the action (\ref{St003}).
We discuss on this in more detail in the next section.

As the simplest  application of the above Hamiltonian analysis let
us calculate the number of field theoretical degrees of freedom of
the dynamical system (\ref{St002}). In this supertwistor
formulation one finds from Eqs.~(\ref{cZtwist}) and (\ref{Ypm})
$(4n+4)$ bosonic and 2 fermionic configuration space variables,
which corresponds to a phase space with $2(4n+4)$ and 4 fermionic
`dimensions'. The system has 5 bosonic first class constraints,
Eqs.~(\ref{PtauY})-(\ref{IPh--Y}), out of a total number of $4n+9$
bosonic constraints (the bosonic components of (\ref{pY}) and
(\ref{psigma3}), (\ref{tau3}), (\ref{Ph++})--(\ref{Ph0})). Thus,
in agreement with Sec. IV, one finds that the
$\Sigma^{({n(n+1)\over 2}|n)}$ supersymmetric string described by the action
(\ref{St002}) possesses $4n-6$ bosonic degrees of freedom.
Likewise, the 2 fermionic constraints of the action (the fermionic
components of (\ref{pY})) reduce to 2 the fermionic degrees of
freedom.

\section{Hamiltonian analysis in terms of
`unnormalized' $\Upsilon^{\pm \Sigma}$ supertwistors}
\label{unnormalized}

As shown in Sec. V, the action (\ref{St002}) may be considered as
a gauge fixed form of the action (\ref{St003}) written in terms of
supertwistors (\ref{Ups}) restricted by only one Lagrangian
constraint (\ref{Con001}). The second constraint (\ref{l+l-00}),
the `normalization' condition that distinguishes among the $Y^{\pm
\Sigma}$ and $\Upsilon^{\pm \Sigma}$ supertwistors, may be
obtained by gauge fixing the direct product of the two scaling
gauge symmetries (\ref{sc+}) and (\ref{sc-}) down to the $SO(1,1)$
worldsheet Lorentz symmetry (\ref{SO(1,1)}) of the action
(\ref{St002}). As a result, one may expect that the Hamiltonian
structure of the model (\ref{St003}) will differ from the one of
the model  (\ref{St002}) by the absence of one second class
constraint  (\ref{N2}) and the presence of one additional first
class constraint replacing (\ref{K0}).

This is indeed the case. An analysis similar to the one carried in
Sec. VI allows one to find the following set of primary
\begin{eqnarray}
\label{pUY}
{\cal P}_{_{\pm\Sigma}} & = & P_{_{\pm
\Sigma}} \mp e_{\sigma}^{\mp \mp} \Omega_{_{\Sigma \Pi}}
\Upsilon^{^{\pm\Pi}} \approx  0 \; ,\\
\label{psU3}
P_{\pm \pm}^{\sigma} & \approx & 0 \; , \\
\label{PtU3} P_{\pm
\pm}^{\tau} & \approx & 0 \; ,   \\
\label{UU1}
{\cal U} \; & := &
\Upsilon^{+\Sigma}\, \Omega_{\Sigma\Pi}
\Upsilon^{-\Pi} \equiv \Upsilon^{+}\, \Omega
\Upsilon^{-}
\approx 0 \; ,
\end{eqnarray}
and secondary constraints
\begin{eqnarray}
\label{Ph++U}
\Phi_{++} &=& \partial_{\sigma} \Upsilon^-
\Omega \Upsilon^- - e^{--}_{\sigma} (\Upsilon^+C\Upsilon^-)^2
\approx 0 \; ,
\\
\label{Ph--U}
\Phi_{--} & =& \partial_{\sigma} \Upsilon^+
\Omega \Upsilon^+ - e^{++}_{\sigma} (\Upsilon^+C\Upsilon^-)^2
\approx 0 \; ,
\\
\label{Ph0U} \Phi^{(0)} &=& \partial_{\sigma} \Upsilon^+ \Omega
\Upsilon^- - \Upsilon^+ \Omega \partial_{\sigma} \Upsilon^-
\approx 0 \; , \qquad
\end{eqnarray}
that restrict the phase space variables
\begin{equation}\label{cZU}
\widetilde{\cal Z}^{\tilde{\cal M}}  \equiv \widetilde{\cal
Z}^{\tilde{\cal M}}(\tau, \sigma) =  \left( \Upsilon^{^{\pm
\Sigma}} \, , \, e^{\pm\pm}_\tau \, , \, e^{\pm\pm}_\sigma
\right)\; ,
\end{equation}
\begin{equation}\label{cPU}
\widetilde{P}_{\tilde{\cal M}}
= (P_{_{\pm\Sigma}}\,,\, P_{\pm\pm}^{\tau} \,, \,
P_{\pm\pm}^{\sigma}) = {\partial {\cal L} \over
\partial
(\partial_\tau \widetilde{\cal Z}^{\tilde{\cal M}})}\; .
\end{equation}

The set (\ref{pUY})--(\ref{Ph0U}) contains $6$ first class
constraints (versus five first class constraints
(\ref{PtauY})--(\ref{IPh--Y}) in the (\ref{St002}) system), namely
\begin{eqnarray}\label{PtauU}
& P_{\pm\pm}^\tau \approx 0 \; , \qquad \\
\label{P1++U} &
2e^{++}_\sigma P_{++}^\sigma-
\Upsilon^{-\Sigma}\mathcal{P}_{-\Sigma}
 \approx 0 \; , \\
\label{P1--U} &
2e^{--}_\sigma P_{--}^\sigma-
\Upsilon^{+\Sigma}\mathcal{P}_{+\Sigma}
 \approx 0 \; , \\
 & \tilde{\Phi}_{++}  =\Phi_{++}
 + \frac{2e^{--}_\sigma {\cal B}^{(0)}}
{(\Upsilon^+C\Upsilon^-)^2}\mathcal{U}  -
 \partial_\sigma P_{++}^\sigma
\nonumber \\
& -  \frac{{\cal B}^{(0)}}{(\Upsilon^+C\Upsilon^-)^2}\Upsilon^{-\Sigma}
 \mathcal{P}_{+\Sigma}
 + (\Upsilon^+C\Upsilon^-)\Upsilon^- C \Omega \mathcal{P}_+ \nonumber \\
 &- \frac1{e_\sigma^{++}} \left[\partial_\sigma
\Upsilon^{-\Sigma} \mathcal{P}_{-\Sigma} +
 \frac{e_\sigma^{--} {\cal B}^{(0)}}{(\Upsilon^+C\Upsilon^-)^2}
 \Upsilon^{+\Sigma} \mathcal{P}_{-\Sigma} \right. \nonumber\\
 & \left. - e^{--}_\sigma (\Upsilon^+C\Upsilon^-)
\Upsilon^+ C \Omega \mathcal{P}_-\right]
\approx 0 \quad , 
\\
 &\tilde{\Phi}_{--}=\Phi_{--} +
\frac{2e^{++}_\sigma {\cal B}^{(0)}}{(\Upsilon^+C\Upsilon^-)^2}
 \mathcal{U} + \partial_\sigma P_{--}^\sigma
\nonumber \\
& -  \frac{{\cal B}^{(0)}}{(\Upsilon^+C\Upsilon^-)^2}\Upsilon^{+\Sigma}
 \mathcal{P}_{-\Sigma}
 + (\Upsilon^+C\Upsilon^-)\Upsilon^+ C \Omega \mathcal{P}_- \nonumber \\
 &+ \frac1{e_\sigma^{--}} \left[\partial_\sigma \Upsilon^{+\Sigma}
 \mathcal{P}_{+\Sigma}
 -\frac{e_\sigma^{++} {\cal B}^{(0)}}{(\Upsilon^+C\Upsilon^-)^2}
 \Upsilon^{-\Sigma} \mathcal{P}_{+\Sigma} \right. \nonumber\\
 & \left. + e^{++}_\sigma (\Upsilon^+C\Upsilon^-)
 (\Upsilon^- C \Omega \mathcal{P}_+)\right]
   \approx 0 \quad ,
 \end{eqnarray}
where ({\it cf.} (\ref{B0=}))
\begin{eqnarray}
& {\cal B}^{(0)} =\frac{1}{2}
\left({\partial_\sigma\Upsilon^+C\Upsilon^+\,
(\Upsilon^+C\Upsilon^-)\over e^{++}_\sigma} +
{\partial_\sigma \Upsilon^-C\Upsilon^-
\, (\Upsilon^+C\Upsilon^-)
\over e^{--}_\sigma} -  \right. \nonumber \\
& - \left.
\frac{\partial_\sigma\Upsilon^+\Omega\partial_\sigma\Upsilon^-}{e^{++}_\sigma
e^{--}_\sigma} \right)\quad .
\end{eqnarray}

Using Dirac brackets to account for the second class constraints
(\ref{pUY}), where ({\it cf.} (\ref{DB2}))
\begin{equation} \label{DB2U}
 [\Upsilon^{^{\pm \Sigma}}(\sigma), \Upsilon^{^{\pm
\Pi}}(\sigma^{\prime})\}_{_D} = \mp {1 \over 2e^{\mp \mp}_{\sigma}} \,
\Omega^{^{\Sigma \Pi}} \, \delta (\sigma - \sigma^{\prime}) \; ,
\end{equation}
the first class constraints simplify to
\begin{eqnarray}\label{PtauUU}
P_{\pm\pm}^\tau &\approx& 0 \; , \qquad \\
\label{P1++UU} P_{++}^\sigma
 &\approx& 0 \; , \\
\label{P1--UU} P_{--}^\sigma
 &\approx& 0 \; , \\
  \tilde{\Phi}_{++}  &=& \Phi_{++}
 + \frac{2e^{--}_\sigma {\cal B}^{(0)}}
{(\Upsilon^+C\Upsilon^-)^2}\mathcal{U}  \approx 0 \; ,
\\
 \tilde{\Phi}_{--}&=&\Phi_{--} + \frac{2e^{++}_\sigma {\cal B}^{(0)}}
 {(\Upsilon^+C\Upsilon^-)^2} \mathcal{U}
 \approx 0 \; ,
\end{eqnarray}
which clearly corresponds to the set of constraints
(\ref{G0(Y)D})--(\ref{IPh--YD}) of the `normalized' supertwistor
description with the addition of the constraint (\ref{K0}), which
is now `converted' into a first class one due to disappearance of
the normalization constraint  (\ref{N2}).

The remaining two bosonic constraints, Eqs. (\ref{UU1}) and
(\ref{Ph0U}), are second class. Their Dirac bracket
\begin{eqnarray}\label{UPh0}
& [{\cal U}(\sigma)\, , \, \Phi^{(0)} (\sigma^\prime)]_D = &
\nonumber \\
&= (\Upsilon^+C\Upsilon^-)^2 \delta(\sigma - \sigma^\prime)& +
\left({\Phi_{++}\over 2e_\sigma^{++}} +
{\Phi_{--}\over 2e_\sigma^{--}} \right) \delta(\sigma - \sigma^\prime)
 \nonumber \\ && \approx
(\Upsilon^+C\Upsilon^-)^2 \delta(\sigma - \sigma^\prime)
  \end{eqnarray}
is nonvanishing due to the linear independence of the
$\Upsilon^{^{+\Sigma}}$ and  $\Upsilon^{^{-\Sigma}}$ supertwistors
(\ref{Ups}) (coming from the linear independence of their
$\Lambda_\alpha^+$ and $\Lambda_\alpha^-$ components,
$\Lambda_\alpha^+C^{\alpha\beta}\Lambda_\alpha^-\not=0$). For a
further simplification of the Hamiltonian formalism it might be
convenient to make a conversion of this pair of second class
constraints into first class by adding a pair of canonically
conjugate variables, $q(\xi)$ and $P^{(q)}(\xi)$, ($[q(\sigma)\, ,
\, P^{(q)}(\sigma^\prime )]_P= \delta(\sigma -\sigma^\prime)$) to
our phase space.

The above Hamiltonian formalism and its further development can be
applied to quantize the $\Sigma^{(\frac{n(n+1)}2|n)}$ supersymmetric string
model. This should produce a quantum higher spin generalization of
the Green--Schwarz superstring for $n=4,8,16$ and, for $n=32$, an
exactly solvable quantum description of a conformal field theory
carrying, somehow, information about the nonperturbative brane BPS
states of M-theory.

\section{Supersymmetric p--branes in maximal superspace 
$\Sigma^{({n(n+1)\over 2}|n)}$} \label{pbrane}

The model may be generalized to describe higher--dimensional
extended objects (supersymmetric $p$--branes) in $\Sigma^{({n(n+1)\over
2}|n)}$.

The expression of the supersymmetric $p$--brane action in terms of
dimensionful unconstrained bosonic spinors reads
(cf.~(\ref{St01}))
\begin{eqnarray}\label{pbr}
S_p &=&  \int_{W^{p+1}} \; e^{\wedge p}_a \wedge \Pi^{\alpha\beta}
(\Lambda^r_\alpha \rho^a_{rs} \Lambda^s_\beta )
\nonumber \\
& -& (- \alpha^\prime )^p  \int_{W^{p+1}} \; e^{\wedge (p+1)} \;
\mathrm{det}(C^{\alpha\beta} \Lambda^r_\alpha \Lambda^s_\beta)\; ,
\qquad
\end{eqnarray}
where $a=0,1, \ldots , p\;,\;
r=1,\ldots,\tilde{n}(p)\,,\;\alpha=1,\ldots,n\,,$
\begin{equation}\label{e10}
e^{\wedge p}_a \equiv
 {1\over p!} \epsilon_{ab_1\ldots
b_{p}}e^{b_1} \wedge \ldots \wedge e^{b_p} \, ,
\end{equation}
and $e^{\wedge (p+1)}$ is the $W^{p+1}$ volume element
\begin{equation}
\label{e11} e^{\wedge (p+1)} \equiv  {1\over (p+1)!}
\epsilon_{b_1\ldots b_{p+1}}e^{b_1} \wedge \ldots \wedge
e^{b_{p+1}}\;\;.
\end{equation}
In Eq.~(\ref{pbr}), the $(p+1)$ $e^a = d\xi^m e_m^a(\xi)$ are
auxiliary worldvolume vielbein fields, $\xi^m=(\tau, \sigma^1,
\ldots, \sigma^p)$ are the worldvolume  $W^{p+1}$ local
coordinates and $\Lambda^r_\alpha(\xi)$ is a set of $\tilde{n}=
\tilde{n}(p)$ unconstrained auxiliary real bosonic fields with a
`spacetime' spinorial (actually, a $Sp(n)$--vector) index $\alpha
=1,..., n$. The number $\tilde{n}(p)$ of real spinor fields
 $\Lambda^r_\alpha(\xi)$ as well as the meaning of the
 symmetric real matrices
$\rho^a_{rs}$ depend on the worldvolume dimension $d=p+1$. For
$d=2,3,4 \;$(mod $8$), where a Majorana spinor representation
exists, the $\rho^a_{rs}$ are $Spin(1,p)$ gamma--matrices
multiplied by the charge conjugation matrix or sigma matrices,
provided they are symmetric. If not, it is always possible to find
a real symmetric matrix by doubling the index $r$, $\check{r}= rI$
 $(I=1,2)$, as in the case of $d=6$ symplectic Majorana
spinors. For dimensions with only Dirac spinors (like $d=5$)
$\Lambda^r_\alpha \rho^a_{rs} \Lambda^s_\beta$ should be
understood as $\bar{\Lambda}_\alpha \gamma^a \Lambda_\beta +
\bar{\Lambda}_\beta \gamma^a \Lambda_\alpha$, {\it etc.} For
simplicity we present Eq. (\ref{pbr}) and other formulae of this
section for `Majorana dimensions' $d$ with symmetric $C
\gamma$--matrices; the generalization to the other cases is
straightforward, although one should be careful determining the
value of $\tilde{n}(p)$ for a given $d=p+1$. For $p=1$, where the
irreducible Majorana--Weyl spinor is one--dimensional ($Spin(1,1)$
is Abelian), one needs $\Lambda_\alpha^r$ to be in a reducible
Majorana representation in the worldsheet spinor index $r$, i.e.
$\Lambda_\alpha^r= (\Lambda_\alpha^+, \Lambda_\alpha^-)$;
otherwise the second term in (\ref{pbr}) would be zero and the
action would become that of a tensionless 
$\Sigma^{({n(n+1)\over 2}|n)}$ supersymmetric string. Then, the
action (\ref{pbr}) reduces to (\ref{St01}) using (\ref{vielbein}).

The fermionic variation $\delta_f$ of the action (\ref{pbr}),
$\delta_{f} S_p$, comes only from the variation of
$\Pi^{\alpha\beta}$. Let us simplify it by taking
 $\delta_{f}X^{\alpha\beta}= i
\delta_{f}{\theta}^{(\alpha}\,{\theta}^{\beta)}$ ({\it cf.} below
Eq.~(\ref{variation})), so that $i_{\delta_{f}}
\Pi^{\alpha\beta}=0$ and $\delta_{f} \Pi^{\alpha\beta}= - 2i
d\theta^{(\alpha} \delta\theta^{\beta)}$. As $\Pi^{\alpha\beta}$
enters in the action (\ref{pbr}) only through its contraction with
$\Lambda^r_\alpha \gamma^a_{rs} \Lambda^s_\beta $ we find
\begin{eqnarray}\label{vSp}
& \delta_{f}  S_p = -2i \int_{W^{p+1}} \, e^{\wedge p}_a \, \wedge
 \,d\theta^{\alpha} \Lambda^r_\alpha \, \rho^a_{rs} \,
\Lambda^s_\beta  \delta\theta^{\beta}\; .
\end{eqnarray}
Thus only $\tilde{n}(p)$ fermionic variations
 $\delta\theta^{\beta} \Lambda^s_\beta$ out of the $n$ variations
 $\delta\theta^{\beta}$ are effectively involved in
$\delta_{f}  S_p$.

This reflects the presence of $(n-\tilde{n}(p))$
$\kappa$--symmetries in the dynamical system described by the
supersymmetric $p$--brane action (\ref{pbr}). They are defined by
\begin{equation}
\label{kapp1} \delta_{\kappa}X^{\alpha\beta}= i
\delta_{\kappa}{\theta}^{(\alpha}\,{\theta}^{\beta)} \; , \quad
\delta_{\kappa} e^a =0\; ,
\end{equation}
and by the following condition on
$\delta_{\kappa}{\theta}^{\alpha}$,
\begin{equation}
\label{kapp2} \delta_{\kappa}{\theta}^{\alpha} \Lambda_{\alpha}^r
=0  \; , \qquad r=1,\ldots , \tilde{n}(p) \; . \qquad
\end{equation}
This can be solved, using the auxiliary spinor fields $u^{\alpha
J}$ [where now $J=1, \ldots, (n-\tilde{n}(p))$] orthogonal to
 $ \Lambda_{\alpha}^r$, as
\begin{eqnarray}
\label{kapp3} \delta_{\kappa}{\theta}^{\alpha}& = & \,
\kappa_{_J}(\xi) \;  u^{\alpha J}(\xi) \; , \qquad
 u^{\alpha J} (\xi)\; \Lambda_{\alpha}^r (\xi ) = 0  \; , \quad
\\ \nonumber && J=1, \ldots , (n-\tilde{n}(p)) \; ,  \quad
r=1,\ldots , \tilde{n}(p) \; . \qquad
\end{eqnarray}
The $\kappa$--symmetry (\ref{kapp1}), (\ref{kapp3}) implies the
preservation of all but $\tilde{n}(p)$ supersymmetries by the
corresponding $\nu=\frac{n-\tilde{n}(p)}{n}$ BPS state.

For instance, for $p=2$, $n=32$, $Spin(1,2) \approx
SL(2,\mathbb{R})$ and $\tilde{n}=2$. The action (\ref{pbr}) then
describes excitations of a membrane BPS state preserving all but
$2$ supersymmetries, a ${30\over 32}$ BPS state. For $p=5$ and
$\tilde{n}=8$ the action (\ref{pbr}) with $n=32$ describes a
${24\over 32}$ supersymmetric 5--brane model in
$\Sigma^{(528|32)}$. Both the su\-per\-mem\-brane (M2--brane) and
the super--5--brane (M5--brane) are known in the standard $D=11$
superspace, where they correspond to ${16\over 32}$ BPS states.
 It is tempting to speculate that the `usual'
M2 and M5 superbranes are related to the generalized
$\Sigma^{(528|32)}$ supersymmetric 2--brane and 5--brane described
by the action (\ref{vSp}) for $p=2$ and $5$. For instance, they
might be related with some particular solutions to the equations
of motion of the corresponding ${30\over 32}$ and ${24\over 32}$
$\Sigma^{(528|32)}$ models preserving $16$ supersymmetries and/or
with the result of a dimensional reduction of them. For the $p=5$
case a question of a special interest would be the role of the M5
selfdual worldvolume gauge field in the $\Sigma^{(528|32)}$
superspace description (see \cite{JdA00} for a related
discussion).

For $p=3$ and $\tilde{n}=4$ we have a ${28\over 32}$ BPS state, a
BPS 3--brane. Neither the Green--Schwarz superstring nor the
super--3--brane exist in the standard $D=11$ superspace, but a
super--D3--brane does exist in the $D=10$ type IIB superspace, as
the superstring does. As we have already noted, although
$\Sigma^{(528|32)}$ also allows a treatment as an enlarged 
type IIB superspace \cite{M-alg,Bars97}, the $\Sigma^{(528|32)}$ 
supersymmetric $p$--brane action
(\ref{pbr}) involves explicitly the $32\times 32$ matrix
$C^{\alpha\beta}$ which cannot be constructed out of type IIB
matrices in a $SO(1,9)$ Lorentz covariant manner. The same problem
appears with the $\Sigma^{(528|32)}$ supersymmetric 9--brane described by the
$p=9$ version of the $\Sigma^{(528|32)}$ model (\ref{pbr}) with
$\tilde{n}=16$, which corresponds to a ${16\over 32}$ BPS state;
its possible relation with the spacetime filling type IIB
super--D9--brane in the usual D=10 superspace is also quite
unclear.

\section{CONCLUSIONS AND OUTLOOK}\label{conclusions}

We have presented a supersymmetric string model in the `maximal' superspace
$\Sigma^{({n(n+1)\over 2}|n)}$ with additional tensorial central
charge coordinates (for $n>2$). The model possesses $n$ rigid
supersymmetries and $n-2$ local fermionic $\kappa$--symmetries.
This implies that it provides the worldsheet action for the
excitations of a BPS state preserving $(n-2)$ supersymmetries. In
particular, for $n=32$ our model describes a supersymmetric string with $30$
$\kappa$--symmetries in $\Sigma^{(528|32)}$, which corresponds to
a BPS state preserving $30$ out of $32$ supersymmetries. This
model can be treated as a composite of two BPS preons \cite{BPS01}
and is the second (after the $D=11$ Curtright model
\cite{Curtright}) tensionful extended object model in
$\Sigma^{(528|32)}$.

In contrast with the Curtright model \cite{Curtright}, our
supersymmetric string action in the enlarged $D=11$ superspace 
$\Sigma^{(528|32)}$ 
does not involve any
gamma--matrices, but instead makes use of two constrained bosonic
spinor variables, $\lambda_\alpha^+$ and $\lambda_\alpha^-$,
corresponding to the two BPS preons from which the superstring BPS
state is composed. As a result, our model preserves the $Sp(32)$
subgroup of the $GL(32,\mathbb{R})$ automorphism symmetry of the
$D=11$ M--algebra.

Our $\Sigma^{({n(n+1)\over 2}|n)}$ supersymmetric string model can be
treated as a higher spin generalization of the classical
Green--Schwarz superstring. At the same time, the additional
bosonic tensorial coordinate fields of the $n=32$ case might
contain information about topological charges corresponding to the
higher branes of the superstring/M-theory [71].

The $\Sigma^{({n(n+1)\over 2}|n)}$ model may also be formulated in
terms of a pair of constrained worldvolume $OSp(2n|1)$
supertwistors. The transition to the supertwistor formulation is
similar to that for the massless superparticle and the tensionless
$\Sigma^{({n(n+1)\over 2}|n)}$ supersymmetric $p$--branes \cite{BL98,
B02}. In our case, however, the supertwistors are restricted by a
constraint that breaks the generalized superconformal $OSp(64|1)$
symmetry down to a generalization of the super--Poincar\'e group,
${\Sigma}^{(528|32)}\times\!\!\!\!\!\!\supset Sp(32)$. Such a
breaking is characteristic of tensionful models. We note that this
constrained $OSp(2n|1)$ supertwistor framework might also be
useful for massive higher spin theories.

We have developed the Hamiltonian formalism, both in the original
and in the symplectic supertwistor representation, and found that,
while the Hamiltonian analysis in the original formulation
requires the use of the additional auxiliary spinor variables
$u_\alpha^I$ ($I=1,...,(n-2)$) orthogonal to $\lambda_\alpha^\pm$,
the symplectic supertwistor Hamiltonian mechanics can be discussed
in terms of the original phase space variables. Moreover, under
Dirac brackets, supertwistors become selfconjugate variables and
the symplectic structure of the phase space simplifies
considerably. A natural application of the Hamiltonian approach
developed here is the BRST quantization of the
$\Sigma^{({n(n+1)\over 2}|n)}$ superstring model, which might
provide a `higher spin' counterpart of the usual string field
theory.

We have also presented a generalization of our $\Sigma^{({n(n+1)\over 2}|n)}$ 
supersymmetric string model
for supersymmetric $p$--branes in $\Sigma^{({n(n+1)\over 2}|n)}$. They
correspond to BPS states preserving all but $\tilde{n}(p)$ (see
below (\ref{pbr})) supersymmetries, composites of $\tilde{n}(p)$
BPS preons $(\tilde{n}(2)=2\,,\; \tilde{n}(3)=4\,,\;
\tilde{n}(5)=8)$. In particular, the ${\Sigma}^{(528|32)}$
supersymmetric membrane ($p=2$) also corresponds to ${30\over 32}$ a BPS
state.

\bigskip

BPS states preserving $30$ out of $32$ supersymmetries have not
been found yet among the solitonic solutions of the `usual' $D=11$
and $D=10$ dimensional supergravities, and the existence of such
solutions is being discussed at present \cite{Duff03,Hull03}. If
found, it would be interesting to study a possible relation of the
additional tensorial bosonic coordinate functions in our theory
with such hypothetical solitonic solutions. In particular, an
interesting question is to see how the WZ term of the superbrane
in usual superspace is reproduced from pure kinetic--like term in
the action (\ref{St}). If, in contrast, these solutions do not
exist, this could indicate, because of the special role of BPS
preons in the algebraic classification of the M--theory BPS states
\cite{BPS01}, the necessity of a wider geometric framework for a
description of M--theory. In this case the proposed 
$\Sigma^{({n(n+1)\over 2}|n)}$ supersymmetric string
model could provide a part of such an extended framework, unifying
M-theory and higher spin theory ideas.

\bigskip

{\it Acknowledgments}. This work has been partially supported by
the research grant BFM2002-03681 from the Ministerio de Ciencia y
Tecnolog\'{\i}a and from EU FEDER funds, by the grant N 383 of the
Ukrainian State Fund for Fundamental Research and by the INTAS
Research Project N 2000-254. M.P. and O.V. wish to thank the
Ministerio de Educaci\'on, Cultura y Deporte and the Generalitat
Valenciana, respectively, for their FPU and FPI research grants.

\bigskip

{\bf Note added}

Two papers \cite{BZ03,DSm03} have just appeared in the net.
Ref.~\cite{BZ03} considers a spontaneous breaking of the
$OSp(1|32)$ symmetry of the tensionless $\Sigma^{({n(n+1)\over
2}|n)}$ supersymmetric $p$--brane models \cite{ZU} \cite{B02}, and
proposes an open tensionless $\Sigma^{({n(n+1)\over 2}|n)}$ 
supersymmetric string action with an additional
boundary term (or topological term, {\it cf.}~\cite{V01s}). These
topological terms can be treated as describing superparticles
attached to the endpoints of a tensionless $\Sigma^{({n(n+1)\over
2}|n)}$ string (similar to quarks attached at the ends of a
bosonic string or D0--branes at the ends of an open superstring,
{\it cf.}~\cite{BH76, B00}).

Ref.~\cite{DSm03} develops a formalism which looks promising for
studying the relation of the BPS preon conjecture and the present
approach with solitonic solutions of the standard $D=11$
supergravity. The authors of \cite{DSm03} deal with bosonic
Killing spinors $\epsilon^{\alpha I}$, but some of their
observations may be applied to the bosonic spinors
$\lambda_\alpha$ ($\lambda_\alpha^r$) characterizing the BPS
preon(s). The  Killing spinors will be orthogonal to
$\lambda_\alpha$ ($\lambda_\alpha^r$) and thus might be identified
with the auxiliary $u^{\alpha I}$ variables of this paper (see \cite{BdAIPV}).

\bigskip

\appendix{{\bf Appendix A. Breaking of the generalized
superconformal group
$OSp(2n|1)$ down to the generalization $\Sigma^{({n(n+1)\over
2}|n)}\times\!\!\!\!\!\!\!\supset Sp(n)$ of the super--Poincar\'e
group.}}
\renewcommand{\theequation}{A.\arabic{equation}}
\setcounter{equation}0

\bigskip

The $(2n+1) \times (2n+1)$ supermatrices ${\cal
G}_{\Sigma}{}^{\Pi} \in OSp(2n|1)$ preserve the
graded--antisymmetric matrix $\Omega_{\Sigma\Pi} =
-(-1)^{\mathrm{deg}({\Sigma}) \mathrm{deg}({\Pi})} \Omega_{\Pi
\Sigma}$, `orthosymplectic metric',
\begin{eqnarray}
\label{GOmG} {\cal G}_{\Sigma}{}^{\Sigma^\prime}
\Omega_{\Sigma^\prime\Pi^\prime} {\cal G}_{\Pi}{}^{\Pi^\prime}
(-1)^{\mathrm{deg}(\Pi)(\mathrm{deg}(\Pi^\prime) + 1)} =
\Omega_{\Sigma\Pi}\; ,
\end{eqnarray}
the canonical form of which is given by Eq. (\ref{OmLP}). The
grading is defined by
$$(-1)^{\mathrm{deg}(\Sigma)} = \begin{cases}
\, 1 & \mathrm{for} \; \Sigma=1, \dots , 2n \cr -1 & \mathrm{for}
\; \Sigma=2n+1 \;
 \end{cases}
$$
and coincides with $\mathrm{deg}(\pm\Sigma)$ for $Y^{\pm\Sigma}$
(see below Eq. (\ref{canonical})). The fundamental representation
of  $OSp(2n|1)$ acts on supertwistors
\begin{eqnarray} \label{Y}
Y^{\Sigma} = (\mu^{\alpha}, \lambda_{\alpha}, \eta ) \; ,
\end{eqnarray}
with even $\mu^{\alpha}, \lambda_{\alpha}$ and odd $\eta$. Near
the unity,
\begin{eqnarray}
\label{GIXi} {\cal G}_{\Sigma}{}^{\Pi} \sim
{\delta}_{\Sigma}{}^{\Pi} + {\Xi}_{\Sigma}{}^{\Pi}\; ,
\end{eqnarray}
where ${\Xi_{\Sigma}}^\Pi$ is an element of the $osp(2n|1)$
superalgebra. It has the form
\begin{eqnarray}
\label{SA} \Xi_{\Sigma}{}^{\Pi} = \left\{ \begin{matrix}
G_{\alpha}{}^\beta & {K}_{\alpha\beta} & \zeta_\alpha \cr
A^{\alpha\beta} & -  G_{\beta}{}^\alpha & \epsilon^\alpha \cr i
\epsilon^\beta & -i \zeta_\beta & 0 \end{matrix} \right\}\; \quad
,
\end{eqnarray}
where the even $n\times n$ matrix $G_{\alpha}{}^\beta$ is
arbitrary and the even $n\times n$ ${K}_{\alpha\beta}=
{K}_{\beta\alpha}$ and ${A}^{\alpha\beta}= {A}^{\beta\alpha}$
matrices  are symmetric. They define a $gl(n)$ and two $sp(n)$
subalgebras  of  $osp(2n|1)$,
\begin{eqnarray}
\label{Agl} G_{\alpha}{}^\beta \, \in \, gl(n)\; , \quad
A^{\alpha\beta} \, \in \, sp(n)\; , \quad \;{K}_{\alpha\beta} \,
\in \, sp(n)\; . \qquad
\end{eqnarray}

Exploiting the analogy with the matrix representation of the
standard $4$--dimensional conformal algebra $su(2,2|N)$ and the
$4$-dimensional super--Poincar\'e algebra, one can look at the
$gl(n)$ boxes $G$ as a generalization of the $spin(1,D-1)$ and
dilatation algebras ($L_\alpha{}^\beta + \delta_\alpha{}^\beta
D$), at the elements $A^{\alpha\beta}\in sp(n)$ as a
generalization of the translation one, and at
${K}_{\alpha\beta}\in sp(n)$ as a generalization of the special
conformal transformations. Eq. (\ref{SA}) also contains two
fermionic parameters, $\epsilon^\alpha$ and $\zeta_\alpha$, which
can be identified as those of the of `usual' and special conformal
supersymmetries. A specific check is provided by the $n=2$ case,
where $SL(2,\mathbb{R})=Spin(1,2)$, the symmetric spin--tensor
provides an equivalent representation for a $SO(1,2)$ vector, and
the superconformal group is $OSp(2|1)$.

If we now demand in addition that the degenerate matrix
$C_{\Sigma\Pi}$ (Eq. (\ref{CLP})) is preserved,
\begin{eqnarray}
\label{GCG} {\cal G}_{\Sigma}{}^{\Sigma^\prime}
C_{\Sigma^\prime\Pi^\prime} {\cal G}_{\Pi}{}^{\Pi^\prime}
(-1)^{\mathrm{deg}(\Pi) (\mathrm{deg}(\Pi^\prime) +1)} =
C_{\Sigma\Pi}\; ,
\end{eqnarray}
we see that this is satisfied by the $osp(2n|1)$ elements of the
form
\begin{eqnarray}
\label{SAC} \Xi_{\Sigma}{}^{\Pi} = \left\{ \begin{matrix}
S_{\alpha}{}^\beta & 0 & 0 \cr A^{\alpha\beta} & -
S_{\beta}{}^\alpha & \epsilon^\alpha \cr i  \epsilon^\beta & 0 & 0
\end{matrix} \right\} \; \equiv  \Xi_{\Sigma}{}^{\Pi} (S,A,
\epsilon) \quad ,
\end{eqnarray}
where $S_{\alpha}{}^\beta \in sp(n)$,
\begin{eqnarray}
 S^{\alpha\beta}\equiv C^{\alpha\gamma} S_\gamma{}^{\beta}
= S^{\beta \alpha} \; ,
\end{eqnarray}
{\it i.e.} by those of (\ref{SA}) with $K_{\alpha\beta}=0$,
$\zeta_\alpha=0$ and $G_{\alpha}{}^\beta =S_{\alpha}{}^\beta  \in
sp(n)$.
 Thus the condition (\ref{GCG}) not only reduces $GL(n)$ symmetry down to
$Sp(n)$, but also breaks the generalized special conformal
transformations and the superconformal supersymmetry.

The right action of ${\cal G}_{\Sigma}{}^{\Pi}(S,A,\epsilon)$
(Eqs. (\ref{GIXi}), (\ref{SAC})) on the supertwistor (\ref{Y}),
$\delta Y^\Sigma= Y^\Pi \Xi_\Pi{}^\Sigma$, defines the generalized
super--Poincar\'e transformation of the supertwistor components,
\begin{eqnarray}
\label{YS-P} \delta \mu^\alpha &=& \mu^\beta S_\beta{}^\alpha +
\lambda_\beta A^{\beta\alpha} + i \epsilon^\alpha \eta \; , \qquad
\nonumber \\
\delta \lambda_\alpha &=& - S_\alpha {}^\beta \lambda_\beta \; ,
\qquad \delta \eta = \epsilon^\alpha \lambda_\alpha \; .
\end{eqnarray}
These can be reproduced from the following transformations of the
$\Sigma^{({n(n+1)\over 2}|n)}$ coordinates
\begin{eqnarray}
\label{dZ-P} \delta X^{\alpha\beta} &=& A^{\alpha\beta} + i
\theta^{(\alpha}\epsilon^{\beta )} + 2 X^{(\alpha|\gamma}
S_\gamma{}^{|\beta)}
\; , \quad \nonumber \\
 \delta \theta^{\alpha}&=&\epsilon^{\alpha} +  \theta^{\beta} A_\beta{}^\alpha
\;
\end{eqnarray}
using the generalization \cite{BL98} of the Penrose correspondence
relation \cite{Pen} \cite{Ferber} given in Eq. (\ref{mu+}),
\begin{equation}
\label{mu}  \mu^{\alpha} = X^{\alpha\beta} \lambda_{\beta}-
{i\over 2} \theta^{\alpha} \theta^{\beta} \lambda_{\beta} \; ,
\quad \eta = \theta^{\alpha} \lambda_{\alpha} \; .
\end{equation}
The transformations (\ref{dZ-P}) of the $\Sigma^{({n(n+1)\over
2}|n)}$ variables are a straightforward generalization of the
super--Poincar\'e transformations of the standard superspace
coordinates. This justifies calling  the resulting supergroup
${\Sigma}^{({n(n+1)\over 2}|n)} \times\!\!\!\!\!\!\supset Sp(n)$ a
generalization of the  super--Poincar\'e group.

Going back to $osp(2n|1)$, let us note that the generalized
special superconformal transformations $(K_{\alpha\beta},
\zeta_\alpha)$ act on the supertwistor components by
\begin{eqnarray}
\label{YS-K} \delta \mu^\alpha = 0 \; , \quad \delta
\lambda_\alpha = \mu^\beta K_{\beta\alpha} - i \eta \zeta_\alpha
\; , \quad \delta \eta =  \mu^\beta  \zeta_\beta \; . \quad
\end{eqnarray}
Using Eq. (\ref{mu}) one may find from (\ref{YS-K}) the
generalized special superconformal transformations of the
${\Sigma}^{({n(n+1)\over 2}|n)}$ coordinates
\begin{eqnarray}
\label{dZ-K} \delta X^{\alpha\beta}\! &=& i \theta^{(\alpha}
X^{\beta)\gamma}\zeta_\gamma - (XKX)^{\alpha\beta}\; ,
\nonumber \\
 \delta \theta^{\alpha}\; &=& X^{\alpha\beta}\zeta_\beta - {i\over 2}
(\theta\zeta)\, \theta^{\alpha} - (\theta KX)^\alpha \; .
\end{eqnarray}

Note that (\ref{dZ-P}) follows as well  from a nonlinear
realization of the generalized super--Poincar\'e group
${\Sigma}^{({n(n+1)\over 2}|n)} \times\!\!\!\!\!\!\supset Sp(n)$
on the ${\Sigma}^{({n(n+1)\over 2}|n)}$ coset, {\it i.e.} from the
left  action of ${\cal G}_{\Sigma}{}^{\Pi}(S, A, \epsilon)  \sim
{\delta}_{\Sigma}{}^{\Pi} + {\Sigma}_{\Sigma}{}^{\Pi}(S, A,
\epsilon) $ (\ref{SAC}) on ${\cal K}_{\Sigma}{}^{\Pi}(X, \theta )
\sim {\delta}_{\Sigma}{}^{\Pi} + K_{\Sigma}{}^{\Pi}(X, \theta)$
with
\begin{eqnarray}
\label{SAX} K_{\Sigma}{}^{\Pi}(X, \theta)= \left\{ \begin{matrix}
0 & 0 & 0 \cr X^{\alpha\beta} &   0 & \theta^\alpha \cr i
\theta^\beta & 0 & 0 \end{matrix} \right\} \quad \; .
\end{eqnarray}
Indeed, the infinitesimal form of
\begin{equation}\label{GK=}
{\cal G}_{\Sigma}{}^{\Pi}(S, A, \epsilon) {\cal
K}_{\Sigma}{}^{\Pi}(X, \theta)= {\cal K}_{\Sigma}{}^{\Pi}(X^\prime
, \theta^\prime) {\cal G}_{\Sigma}{}^{\Pi}(A, 0, 0)\;
\end{equation}
reads
\begin{eqnarray}\label{GKi=}
{K}(\delta X, \delta \theta) &=& {\Xi}(0, A, \epsilon) + {\Xi}(0,
A, \epsilon) {K}(X, \theta)
\nonumber \\
&+& [ {\Xi}(S, 0, 0)\, , \, K(X, \theta)]\;
\end{eqnarray}
and reproduces the generalized super--Poincar\'e transformations
(\ref{dZ-P}).

\bigskip

\appendix{{\large\bf Appendix B. Some technical details.}}
\renewcommand{\theequation}{B.\arabic{equation}}
\setcounter{equation}0

\subsection*{General solution of Eqs.~(\ref{dtP})--(\ref{dtP0}) for
the Lagrange multipliers (Eqs.~(\ref{Lab=})--(\ref{l0=}))}

\begin{eqnarray}
\label{Lab=A}
L^{\alpha\beta} &= & b_{IJ}u^{\alpha I} u^{\beta J} + \nonumber \\
&+& {e^{++}_{\tau}\over e^{++}_{\sigma}}
\left[ e^{++}_{\sigma} \lambda^{-\alpha}\lambda^{-\beta} +
2 \left(\lambda^-_\gamma \Pi_{\sigma}^{\gamma(\alpha} \lambda^{+\beta)}
\right.\right. \nonumber \\
 &-& \left.\left.(\lambda^-\Pi_\sigma
\lambda^+)\lambda^{-(\alpha}\lambda^{+\beta)}+(\lambda^-\Pi_\sigma
\lambda^-)\lambda^{+(\alpha}\lambda^{+\beta)}
  \right)\right]\, +
\nonumber \\ &+&
{e^{--}_{\tau} \over e^{--}_{\sigma}}
\left[e^{--}_{\sigma}\lambda^{+\alpha}\lambda^{+\beta} - 2
\left(\lambda^+_\gamma \Pi_{\sigma}^{\gamma(\alpha} \lambda^{-\beta)}\right.
\right. \nonumber \\
 &-& \left.\left.(\lambda^+\Pi_\sigma
\lambda^+)\lambda^{-(\alpha}\lambda^{-\beta)}+(\lambda^+\Pi_\sigma
\lambda^-)\lambda^{+(\alpha}\lambda^{-\beta)}
  \right)\right]\, , \nonumber \\ {} \\
\label{xi=A} \xi^{\alpha} &=& \kappa_I \, u^{\alpha I} +
\frac{e^{++}_{\tau}}{e^{++}_{\sigma}}
(\partial_{\sigma} \theta \lambda^-) \lambda^{+ \alpha} -
\frac{e^{--}_{\tau}}{e^{--}_{\sigma}} (\partial_{\sigma} \theta
\lambda^+) \lambda^{- \alpha}
\; ,
\nonumber \\ {}
\end{eqnarray}
\begin{eqnarray}
 \label{l+=A} l^+_{\alpha} &= & \omega^{(0)}\lambda_{\alpha}^{+} +
{e^{--}_{\tau} \over e^{--}_{\sigma}}
\left(\partial_{\sigma}\lambda^+_\alpha - \Omega^{(0)}_\sigma
\lambda^+_\alpha\right) + \nonumber \\ &+& { e^{--}_{\tau} \over
2e^{--}_{\sigma}e^{--}_{\sigma}} \left(-e^{--}_{\sigma}
\Omega^{++}_{\sigma} - e^{++}_{\sigma} \Omega^{--}_{\sigma} +
i\partial_{\sigma}\theta \lambda^+\partial_{\sigma}\theta
\lambda^- - \right. \nonumber \\ &&  \left.  \qquad -
\Pi^{\alpha\beta}_\sigma (\partial_{\sigma}\lambda^+_{\alpha}
\lambda^-_{\beta} -  \lambda^+_{\alpha}
\partial_{\sigma}\lambda^-_{\beta}) \right)\lambda^-_\alpha +
\nonumber \\ &+& { e^{++}_{\tau} \over
2e^{++}_{\sigma}e^{--}_{\sigma}} \left(e^{--}_{\sigma}
\Omega^{++}_{\sigma} + e^{++}_{\sigma} \Omega^{--}_{\sigma} +
i\partial_{\sigma}\theta \lambda^+\partial_{\sigma}\theta
\lambda^- + \right. \nonumber \\ &&  \left.  \qquad +
\Pi^{\alpha\beta}_\sigma (\partial_{\sigma}\lambda^+_{\alpha}
\lambda^-_{\beta} -  \lambda^+_{\alpha}
\partial_{\sigma}\lambda^-_{\beta}) \right)\lambda^-_\alpha \; ,
\end{eqnarray}
\begin{eqnarray}
\label{l-=A}  l^-_{\alpha} &=& -\omega^{(0)}\lambda_{\alpha}^{-} +
{e^{++}_{\tau} \over e^{++}_{\sigma}}
\left(\partial_{\sigma}\lambda^-_\alpha + \Omega^{(0)}_\sigma
\lambda^-_\alpha\right) + \nonumber \\
&+& { e^{--}_{\tau} \over 2e^{++}_{\sigma}e^{--}_{\sigma}} \left(
-e^{--}_{\sigma} \Omega^{++}_{\sigma} - e^{++}_{\sigma}
\Omega^{--}_{\sigma} + i\partial_{\sigma}\theta
\lambda^+\partial_{\sigma}\theta
\lambda^- - \right. \nonumber \\
&&  \left.  \qquad - \Pi^{\alpha\beta}_\sigma
(\partial_{\sigma}\lambda^+_{\alpha} \lambda^-_{\beta} -
\lambda^+_{\alpha}
\partial_{\sigma}\lambda^-_{\beta}) \right)\lambda^+_\alpha + \nonumber \\
&+& {e^{++}_{\tau} \over 2e^{++}_{\sigma}e^{++}_{\sigma}}
\left(e^{--}_{\sigma} \Omega^{++}_{\sigma} + e^{++}_{\sigma}
\Omega^{--}_{\sigma} + i\partial_{\sigma}\theta
\lambda^+\partial_{\sigma}\theta \lambda^- + \right. \nonumber \\
&&  \left.  \qquad + \Pi^{\alpha\beta}_\sigma
(\partial_{\sigma}\lambda^+_{\alpha} \lambda^-_{\beta} -
\lambda^+_{\alpha} \partial_{\sigma}\lambda^-_{\beta})
\right)\lambda^+_\alpha \; ,
\end{eqnarray}
\begin{eqnarray}
\label{Lpmpm=A} &  L^{\pm\pm}= \partial_{\sigma}e^{\pm \pm}_{\tau}
+ 2e^{\pm \pm}_\tau  \Omega^{(0)}_\sigma \pm 2e^{\pm \pm}_{\sigma}
\omega^{(0)} \; ,
\end{eqnarray}
\begin{eqnarray}
\label{L(n)=A} &  L^{(n)}= - 4 \mathrm{det}(e_m^a) \equiv -2
(e_\tau^{--}e_\sigma^{++} - e_\tau^{++}e_\sigma^{--} ) \; ,
\end{eqnarray}
\begin{eqnarray}
\label{l0=A}&  l^{(0)}= 0\; ,
\end{eqnarray}
where, $\Omega_\sigma^{\pm\pm}$ and $\Omega_\sigma^{(0)}$ are defined in
(\ref{Ompm=}), (\ref{Om0=}), namely
\begin{eqnarray}\label{Ompm=A}
\Omega_\sigma^{++} &:=& \partial_{\sigma} \lambda^+C \lambda^+\; ,
\qquad \Omega_\sigma^{--} :=\partial_{\sigma} \lambda^-C \lambda^-
\\ \label{Om0=A} \Omega_\sigma^{(0)} &:=&
\frac{1}{2}(\partial_{\sigma} \lambda^+C \lambda^- -
\lambda^+C \partial_{\sigma} \lambda^-) \; .
\end{eqnarray}

\subsection*{Dirac brackets of the second class constraints
(\ref{K0})--(\ref{Ph0II})}

\begin{eqnarray} \label{DB1}
&& [\Phi^{(0)}(\sigma) , {\cal U}(\sigma^\prime)]_{_D} = \nonumber \\
&& = -\frac{1}{2}
\left( \frac{\partial_\sigma Y^{+\Sigma}(\sigma)
\Omega_{\Sigma \Pi} Y^{+ \Pi}(\sigma)}{e^{++}_{\sigma}(\sigma)} + \right.
\nonumber \\
&& \left.
\qquad +\frac{\partial_\sigma Y^{-\Sigma}(\sigma)
\Omega_{\Sigma \Pi} Y^{- \Pi}(\sigma)}{e^{--}_\sigma(\sigma)} \right)
\delta(\sigma-\sigma^\prime) = \nonumber \\
&& = -\frac{1}{2} \left( \frac{\Phi_{++}(\sigma)}{e^{++}_\sigma(\sigma)}
+\frac{\Phi_{--}(\sigma)}{e^{--}_\sigma(\sigma)} +2 \right)
\delta(\sigma-\sigma^\prime) \approx
\nonumber \\
&& \approx -\delta(\sigma-\sigma^\prime) \; ,
\end{eqnarray}
\begin{eqnarray} \label{DBPh0N}
&& [\Phi^{(0)}(\sigma) , {\cal N}(\sigma^\prime)]_{_D} = \nonumber \\
&&= -\frac{1}{2}
\left( \frac{\partial_\sigma Y^{+\Sigma}(\sigma)
C_{\Sigma \Pi} Y^{+ \Pi}(\sigma)}{e^{++}_{\sigma}(\sigma)} + \right.
\nonumber \\
&& \left.
\qquad +\frac{\partial_\sigma Y^{-\Sigma}(\sigma)
C_{\Sigma \Pi} Y^{- \Pi}(\sigma)}{e^{--}_\sigma(\sigma)} \right)
\delta(\sigma-\sigma^\prime) = \nonumber \\
&& = -\frac{1}{2}
\left( \frac{A_{\sigma}^{++}(\sigma)}{e^{++}_\sigma(\sigma)} +
\frac{A_{\sigma}^{--}(\sigma)}{e^{--}_\sigma(\sigma)} \right)
\delta(\sigma-\sigma^\prime) \; ,
\end{eqnarray}
\begin{eqnarray} \label{DB3}
&& [K^{(0)}(\sigma) , {\cal U}(\sigma^\prime)]_{_D} =
\nonumber \\ && \qquad  =
Y^{+\Sigma}(\sigma)\Omega_{\Sigma \Pi} Y^{-\Pi}(\sigma) \
\delta(\sigma-\sigma^\prime) =
\nonumber \\ && \qquad = {\cal U} \ \delta(\sigma-\sigma^\prime) \approx 0 \; ,
\end{eqnarray}
\begin{eqnarray} \label{DB4}
&& [K^{(0)}(\sigma) , {\cal N}(\sigma^\prime)]_{_D} =
\nonumber \\ && \qquad  =
Y^{+\Sigma}(\sigma)C_{\Sigma \Pi} Y^{-\Pi}(\sigma) \
\delta(\sigma-\sigma^\prime) =
\nonumber \\ && \qquad = ({\cal N} +1) \ \delta(\sigma-\sigma^\prime) \approx
\delta(\sigma-\sigma^\prime) \; ,
\end{eqnarray}
\begin{eqnarray} \label{DB5}
&& [K^{(0)}(\sigma) , \Phi^{(0)}(\sigma^\prime)]_{_D} =
\nonumber \\ &&
=\frac{1}{2} \left( \partial_\sigma Y^{+\Sigma}(\sigma)\Omega_{\Sigma \Pi}
Y^{-\Pi}(\sigma) - \right.
\nonumber \\ && \left. \qquad - Y^{+\Sigma}(\sigma)\Omega_{\Sigma \Pi}
\partial_\sigma Y^{-\Pi}(\sigma) \right)
\delta(\sigma-\sigma^\prime) =
\nonumber \\ && =\frac{1}{2} \Phi^{(0)} \ \delta(\sigma-\sigma^\prime)
\approx 0 \; .
\end{eqnarray}

\end{document}